\pdfoutput=1
\documentclass[12pt]{iopart}
\usepackage{iopams}
\usepackage{graphicx}
\usepackage{setstack}

\newcommand{\rmf}{\mathrm{f}}
\newcommand{\hrho}{\hat{\rho}}
\newcommand{\hbrho}{\hat{\brho}}

\begin{document}
\title{Singularities in large deviation functions}

\author{Yongjoo Baek and Yariv Kafri}
\address{Department of Physics, Technion, Haifa 32000, Israel}
\eads{yongjoo.baek@physics.technion.ac.il and kafri@physics.technion.ac.il}

\begin{abstract}
Large deviation functions of configurations exhibit very different behaviors in and out of thermal equilibrium. In particular, they exhibit singularities in a broad range of non-equilibrium models, which are absent in equilibrium. These singularities were first identified in finite-dimensional systems in the weak-noise limit. Recent studies have shown that they are also present in driven diffusive systems with an infinite-dimensional configuration space. This short review describes singularities appearing in both types of systems under a unified framework, presenting a classification of singularities into two broad categories. The types of singularities which were identified for finite-dimensional cases are compared to those found in driven diffusive systems.
\end{abstract}

\maketitle

\section{Introduction}

There are many physical problems in which one is interested in understanding the statistics of rare fluctuations. These are often encoded in a {\it large deviation function} (LDF), or a {\it rate function}~\cite{Varadhan1984,Ellis1985,Dembo1998,Touchette2009}. For this reason, LDFs have been studied in the context of fluctuation theorems~\cite{Lebowitz1999,Saito2011}, dynamics of relaxation processes~\cite{Maes2011, Lee2013}, current statistics~\cite{Bodineau2004,Harris2005,Lecomte2010,Gorissen2012,Akkermans2013}, systems with long-range interactions~\cite{Barre2005}, diffusion of tracer particles~\cite{Krapivsky2014}, population dynamics~\cite{Kamenev2008}, condensation phenomena~\cite{Merhav2010,SzavitsNossan2014,Zannetti2014}, non-equilibrium phase transitions~\cite{Cohen2012}, quantum work statistics~\cite{Gambassi2012}, and many more. Furthermore, efficient numerical methods for calculating LDFs have also been devised~\cite{Giardina2006,Lecomte2006,Giardina2006B,Bunin2012a,Nemoto2014}.

One class of LDFs which have attracted much attention are those associated with density profiles in spatially extended {\it driven diffusive systems} out of thermal equilibrium~\cite{Bertini2001,Derrida2001,Derrida2002,Derrida2003,Enaud2004,Bertini2005b,Derrida2007,Bertini2014}. In such systems the LDFs are a direct analogue of free energies in equilibrium systems. The latter, in driven diffusive systems, can easily be shown (see discussions in \ref{ssec:db}) to be local and smooth functions of density profiles. In contrast, the former have distinct properties out of equilibrium. First, the LDFs are generally {\it non-local} functions of density profiles, as might be expected from the presence of long-range correlations in bulk-conserving non-equilibrium systems~\cite{Machta1980,Spohn1983,Dorfman1994,Bunin2013b}. Moreover, in some models the LDFs are non-differentiable \cite{Bertini2010,Bunin2012b,Bunin2013a,Aminov2014}. These singular behaviors can be considered as extensions of similar behaviors in systems with a finite-dimensional configuration space, first observed by Graham and T\'{e}l~\cite{Graham1984a,Graham1984b} in the {\it weak-noise limit}, to systems with an infinite-dimensional configuration space. In the latter the weak-noise limit arises naturally due to the macroscopic system size~\cite{Spohn1991}.

The aim of this short review is to give an overview of the known results for finite-dimensional systems and those for infinite-dimensional systems within the same framework. As should become clear to the reader, there are singular behaviors in finite-dimensional systems which have not been identified so far in infinite-dimensional systems. Moreover, the LDFs in infinite-dimensional systems can exhibit behaviors that are richer than the finite-dimensional counterparts. This leaves much room for further studies.

To this end, we provide a classification of the known singular behaviors of finite-dimensional systems, so that the recently found singularities in infinite-dimensional systems can be put in the context of the former. We also pay special attention to the duality between a weak-noise system and a Hamiltonian system, whose implications on singular LDFs were discussed in great details for finite-dimensional cases but less so for infinite-dimensional ones. This sheds new light on the physical meaning of the ``order parameter'' that appears in the descriptions~\cite{Bunin2013a,Aminov2014} of singular structures observed in infinite-dimensional systems.

This review deals only with theoretical aspects of singularities shown by LDFs of density profiles or, more generally speaking, system configurations. For a review of experimental aspects of this subject, see \cite{Luchinsky1998}. We note that LDFs of other variables also exhibit singularities, whose properties and origins are generally different from those of the singularities discussed in this review. For example, LDFs of the steady-state current may exhibit singularities associated with dynamical phase transitions, even though the LDFs for density profiles remain smooth. An interested reader is referred to \cite{Bodineau2005,Bertini2005a,Bertini2006,Hurtado2011}.

This review is organized as follows. In \sref{sec:formalism}, we provide the general formalism for finite-dimensional systems in the weak-noise limit, describing how their LDFs can be obtained from the associated Hamiltonian structure. Based on this formalism, in \sref{sec:fin_dim} we discuss different scenarios that give rise to singularities of these LDFs, listing relevant studies for each scenario. Section~\ref{sec:inf_dim} gives a review of singular behaviors found in driven diffusive systems, emphasizing how they fit into the general framework set up by finite-dimensional cases. A summary and an outlook for future works are given in \sref{sec:summary}.

\section{General formalism} \label{sec:formalism}

\subsection{The Hamiltonian structure of weak-noise systems}
\subsubsection{Equations of motion}

We consider a system whose configuration is represented by an $n$-component vector $\brho \equiv (\rho_1, \rho_2, \ldots, \rho_n)$. The system evolves in time $t$ according to the Langevin equation
\begin{equation} \label{eq:langevin}
\dot{\brho}(t) = \bi{F}[\brho(t)] + G[\brho(t)] \bfeta(t),
\end{equation}
where $\bi{F} \equiv (F_1, F_2, \ldots, F_n)$ denotes a deterministic drift force, $\bfeta \equiv (\eta_1, \eta_2, \ldots, \eta_m)$ an $m$-component noise of the environment, and $G$ an $n$-by-$m$ matrix. While for equilibrium systems $\bi{F}$ and $G$ are related by a fluctuation--dissipation relation, here no such relation is assumed; at the moment, we only assume that all elements of $\bi{F}$ and $G$ are smooth (i.e. infinitely differentiable) functions. In addition, we assume $\bfeta$ to be a Gaussian white noise with
\begin{eqnarray} \label{eq:noise}
\langle \eta_\mu(t) \rangle = 0, \quad \langle \eta_\mu(t) \eta_\nu(t') \rangle = N^{-1} \delta_{\mu\nu} \delta(t-t'),
\end{eqnarray}
so that the noise amplitude is proportional to $N^{-1/2}$. Our interest lies in the large $N$ limit.

In the noiseless case, the system evolves through the deterministic dynamics
\begin{equation} \label{eq:noiseless}
\dot{\brho}(t) = \bi{F}[\brho(t)].
\end{equation}
We are interested in cases when the solutions of this equation approach the limit sets (e.g. attractors, repellers, saddles) denoted by $\Gamma_1, \Gamma_2, \ldots$ as $t \to \pm \infty$. The attractors among these limit sets correspond to the (meta)stable states of the system. If the noise is very weak but nonzero ($N^{-1} \ll 1$), then we expect a steady-state distribution satisfying the large deviation principle
\begin{equation} \label{eq:ld_principle}
P_{\rm s}(\brho) \sim \exp [-N\phi(\brho)],
\end{equation}
with local minima of the LDF $\phi$ coinciding with the attractors of \eref{eq:noiseless}.

A rigorous proof of \eref{eq:ld_principle} is beyond the scope of this review, and the interested reader is referred to \cite{Freidlin2012}. Here we give a heuristic description of how the LDF $\phi$ is obtained. We start by writing the propagator $P(\brho_\rmf,t_\rmf|\brho_\rmi,t_\rmi)$ in a path integral form. The propagator is obtained by summing the probability of every trajectory $\brho(t)$ satisfying the Langevin equation \eref{eq:langevin} and the boundary conditions $\brho(t_\rmi)=\brho_\rmi$ and $\brho(t_\rmf)=\brho_\rmf$, as expressed by
\begin{equation} \label{eq:path_int}
P(\brho_\rmf,t_\rmf|\brho_\rmi,t_\rmi) = \left\langle \int_{\brho(t_\rmi)=\brho_\rmi}^{\brho(t_\rmf)=\brho_\rmf} {\cal D}\brho \, {\cal J}(\brho) \delta (\dot{\brho} - \bi{F} - G \bfeta) \right\rangle_{\bfeta}.
\end{equation}
Here $\langle \cdot \rangle_{\bfeta}$ denotes the average with respect to the distribution of $\bfeta$, and ${\cal J}(\brho)$ the Jacobian for changing the variable of integration from $\dot{\brho} - \bi{F} - G\bfeta$ to $\brho$. While ${\cal J}(\brho)$ depends on the discretization (e.g. It\={o} or Stratonovich) of the Langevin equation \eref{eq:langevin}, it always has the form $\exp\,[O(1)]$ (e.g. ${\cal J}(\brho) = 1$ for the It\={o} discretization)~\cite{ZinnJustin2002}. Thus, in the large $N$ limit, the differences between the It\={o} and Stratonovich schemes are irrelevant to the LDF given by \eref{eq:ld_principle}. We can therefore neglect ${\cal J}(\brho)$ and evaluate the average $\langle \cdot \rangle_{\bfeta}$ to obtain
\begin{eqnarray}
\fl P(\brho_\rmf,t_\rmf|\brho_\rmi,t_\rmi)
&= \int_{\brho(t_\rmi)=\brho_\rmi}^{\brho(t_\rmf)=\brho_\rmf}{\cal D}[\brho,\bfeta] \, \delta(\dot{\brho} - \bi{F} - G \bfeta) \exp \left( -N \int_{t_\rmi}^{t_\rmf} \rmd t \, \frac{\bfeta^2}{2} \right) \\
\fl &= \int_{\brho(t_\rmi)=\brho_\rmi}^{\brho(t_\rmf)=\brho_\rmf}{\cal D}[\brho] \, \exp \Bigg[ -N \int_{t_\rmi}^{t_\rmf} \rmd t \, \underbrace{\frac{1}{2}(\dot{\brho}-\bi{F})\cdot\chi^{-1}(\dot{\brho}-\bi{F})}_{\equiv L(\brho,\dot{\brho})} \Bigg]. \label{eq:lagrangian}
\end{eqnarray}
Here the $n$-by-$n$ matrix $\chi \equiv G^T G$ is assumed to be a positive-definite matrix, which is thus invertible. In the large $N$ limit, we can apply the saddle-point method to obtain
\begin{equation} \label{eq:propagator}
P(\brho_\rmf,t_\rmf|\brho_\rmi,t_\rmi) \sim \exp [ -N S(\brho_\rmf,t_\rmf;\brho_\rmi,t_\rmi) ],
\end{equation}
where we introduced a function defined by
\begin{equation} \label{eq:action_lagrangian}
S(\brho_\rmf,t_\rmf;\brho_\rmi,t_\rmi) \equiv \min_{\brho(t)} \int_{t_\rmi}^{t_\rmf} \rmd t \,L(\brho,\dot{\brho}),
\end{equation}
with the minimum taken over all trajectories $\brho(t)$ satisfying the boundary conditions $\brho(t_\rmi)=\brho_\rmi$ and $\brho(t_\rmf)=\brho_\rmf$. We observe that \eref{eq:action_lagrangian} has the form of a least action principle whose Lagrangian is $L(\brho,\dot{\brho})$. Introducing the momenta $\hbrho \equiv (\hrho_1, \hrho_2, \ldots, \hrho_n)$, whose components are defined by $\hrho_\mu \equiv \partial L/ \partial \dot{\rho}_\mu$, we can rewrite the least action principle as
\begin{equation} \label{eq:action}
S(\brho_\rmf,t_\rmf;\brho_\rmi,t_\rmi) = \min_{\brho(t),\hbrho(t)} \int_{t_\rmi}^{t_\rmf} \rmd t \, \left[ \hbrho \cdot \dot{\brho} - H(\brho,\hbrho) \right],
\end{equation}
where the Hamiltonian is given by the Legendre transform
\begin{equation} \label{eq:hamiltonian}
H(\brho,\hbrho) = \hbrho \cdot \dot{\brho} - L(\brho,\dot{\brho}) = \hbrho \cdot \bi{F}(\brho) + \frac{1}{2}\hbrho\cdot \chi(\brho) \hbrho.
\end{equation}
This conversion to a Hamiltonian picture can also be done within the path integral formulation using a Hubbard--Stratonovich transformation or the Martin--Siggia--Rose (MSR) procedure~\cite{Martin1973,DeDominicis1976,Janssen1976,DeDominicis1978}. See \ref{app:hamiltonian} for details about the latter.

The least action principle of \eref{eq:action} indicates that the propagator is dominated by {\it locally minimizing trajectories} which satisfy the following Hamilton equations
\numparts
\begin{eqnarray}
\dot{\brho} = \frac{\partial H}{\partial \hbrho} = \bi{F}(\brho) + \chi(\brho) \hbrho,\label{eq:hamilton_coords}\\
\dot{\hbrho} = -\frac{\partial H}{\partial \brho} = -\sum_{\mu = 1}^n \hrho_\mu \frac{\partial F_\mu}{\partial \brho} - \frac{1}{2} \sum_{\mu = 1}^{n}\sum_{\nu = 1}^{n} \hrho_\mu \hrho_\nu \frac{\partial \chi_{\mu\nu}}{\partial \brho} \label{eq:hamilton_momenta}.
\end{eqnarray}
\endnumparts
We note that if $\hbrho = 0$, these equations describe the noiseless dynamics of \eref{eq:noiseless}. Thus, the momenta $\hbrho$ represents the effects of noise on the minimizing trajectories. We also note that these equations by themselves describe trajectories {\it extremizing} the action rather than minimizing it. Thus, when referring to the general solutions of \eref{eq:hamilton_coords}, \eref{eq:hamilton_momenta}, we shall use the term {\it extremizing trajectories}, from which locally minimizing trajectories must be distinguished.

So far, we have shown that there is a correspondence between the weak-noise limit of the Langevin equation \eref{eq:langevin} and a Hamiltonian system governed by \eref{eq:hamilton_coords} and \eref{eq:hamilton_momenta}. We next turn to the structure of the solutions of these Hamilton equations.

\subsubsection{Structure of solutions} \label{sssec:structure}

\begin{figure}
\begin{center}
\includegraphics[width=0.7\textwidth]{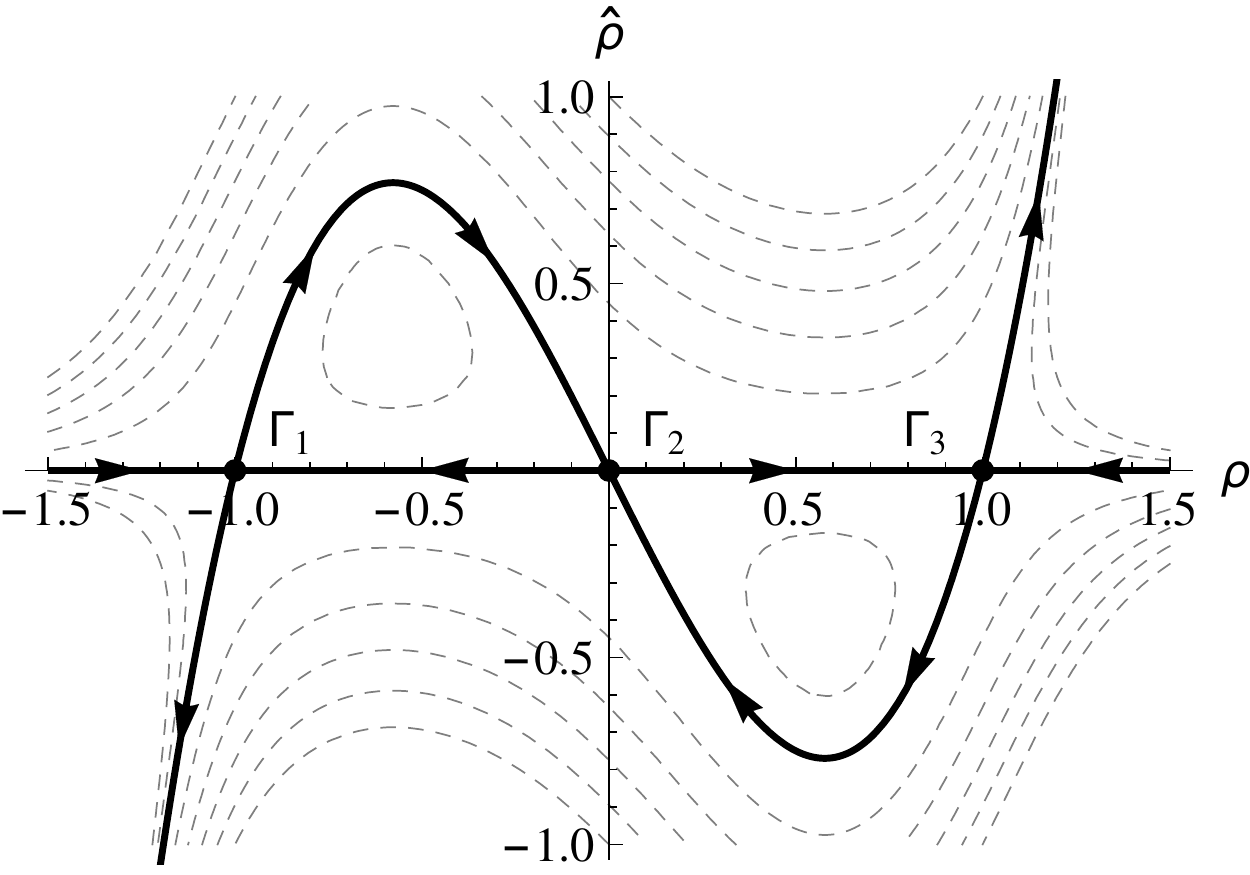}
\end{center}
\caption{\label{fig:ex1_phase_space} Phase-space trajectories of the Hamiltonian system satisfying \eref{eq:example1}. Dashed (thick solid) lines represent the trajectories of non-zero (zero) energy. Directions of motion are indicated by arrows.}
\end{figure}

To make our discussions more intuitive, we consider an example of the Langevin equation \eref{eq:langevin} given by
\begin{equation} \label{eq:example1}
F(\rho) = -V'(\rho) = \rho - \rho^3, \qquad \chi(\rho) = 1,
\end{equation}
which describes an overdamped Brownian motion inside a double-well potential $V(\rho) = \rho^4/4 - \rho^2/2$. Applying \eref{eq:hamiltonian}, the weak-noise limit of this system is described by the Hamiltonian
\begin{equation} \label{eq:ex_hamiltonian}
H(\rho,\hrho) = \hrho \, (\rho-\rho^3) + \frac{\hrho^2}{2},
\end{equation}
whose corresponding equations of motion are obtained from \eref{eq:hamilton_coords} and \eref{eq:hamilton_momenta} by
\begin{equation}\label{eq:ex_hamilton_coords}
\dot{\rho} = \rho - \rho^3 + \hrho, \quad \dot{\hrho} = \hrho \, (3\rho^2 - 1). \\
\end{equation}
Figure~\ref{fig:ex1_phase_space} shows the phase-space trajectories traced by solutions of these equations. We make a few observations on the structure of these solutions, which are generally true for any Hamiltonian system satisfying \eref{eq:hamilton_coords} and \eref{eq:hamilton_momenta}.

\begin{enumerate}
\item The limit sets of the Hamiltonian system contain the limit sets of the noiseless dynamics, because the latter can be obtained from the former by putting $\hbrho = 0$. For example, denoting a point in the phase space by $(\rho,\hrho)$, \eref{eq:ex_hamilton_coords} has three limit sets $\Gamma_1 = \{(-1,0)\}$, $\Gamma_2 = \{(0,0)\}$, and $\Gamma_3 = \{(1,0)\}$, which are shown in \fref{fig:ex1_phase_space}. These limit sets are also the limit sets of the noiseless dynamics represented by the trajectories on the $\hrho = 0$ line, for which $\Gamma_1$ and $\Gamma_3$ are attractors and $\Gamma_2$ is a repeller. Note that while all the limit sets mentioned here are points, in general they may be higher-dimensional objects such as periodic orbits or limit cycles.

\item The Hamiltonian system does not have any pure attractors or pure repellers. This is a general property due to Liouville's theorem, which states that the area of the phase space is preserved by Hamiltonian dynamics. As \fref{fig:ex1_phase_space} shows, the attractors $\Gamma_1$ and $\Gamma_3$ (repeller $\Gamma_2$) of the noiseless dynamics are repulsive (attractive) for ``noisy'' trajectories with $\hrho \neq 0$. Thus, the system can escape from any limit set towards a neighboring limit set. Recall that this Hamiltonian dynamics is derived in the $N \to \infty$ limit, which implies that the evolution from one limit set to another requires an infinite time limit ($t_\rmi \to -\infty$ or $t_\rmf \to +\infty$). This allows us to define a directed graph $\cal G$ whose edge set $E(\cal G)$ contains an edge $(\alpha \to \beta)$ if there is such a trajectory from $\Gamma_\alpha$ to $\Gamma_\beta$. We may call $\alpha$ an in-neighbor of $\beta$ and $\beta$ an out-neighbor of $\alpha$, denoting the relationships by $\alpha \in {\cal N}^- (\beta)$ and $\beta \in {\cal N}^+ (\alpha)$, respectively. For example, we can construct a graph with the edge set $E({\cal G}) = \{(1 \to 2), (2 \to 1), (2 \to 3), (3 \to 2)\}$ from the trajectories shown in \fref{fig:ex1_phase_space}. While the noiseless trajectories produce the relations ${\cal N}^+(2) = \{1,3\}$ and ${\cal N}^-(1) = {\cal N}^-(3) = \{2\}$, the noisy trajectories give rise to the inverted relations ${\cal N}^-(2) = \{1,3\}$ and ${\cal N}^+(1) = {\cal N}^+(3) = \{2\}$.

\item Trajectories going through the limit sets of the noiseless dynamics have ``zero energy'' in the sense that the Hamiltonian stays zero along these trajectories (thick solid lines in \fref{fig:ex1_phase_space}). The Hamiltonian of the form \eref{eq:hamiltonian} is not explicitly dependent on time, so all trajectories of the Hamiltonian system are confined to a surface of constant energy. Since the trajectories going through the limit sets have $\hbrho = 0$ at some point, their constant energy must be zero. For example, \fref{fig:ex1_phase_space} shows that the trajectories connecting different limit sets must satisfy $\hrho = 0$ or $\hrho = 2(\rho^3-\rho)$, both implying $H(\rho,\hrho) = 0$ by \eref{eq:ex_hamiltonian}.

\item Trajectories with nonzero energy accumulate an indefinitely large positive action in the limit $t_\rmi \to -\infty$ or $t_\rmf \to +\infty$. Using \eref{eq:lagrangian} and \eref{eq:hamilton_coords}, the action defined in \eref{eq:action_lagrangian} can be rewritten as
\begin{equation} \label{eq:lagrangian_momenta}
S(\brho_\rmf,t_\rmf;\brho_\rmi,t_\rmi) = \min_{\hbrho(t)} \int_{t_\rmi}^{t_\rmf} \rmd t \, \frac{1}{2}\, \hbrho \cdot \chi(\brho) \hbrho,
\end{equation}
where the minimum is taken over trajectories of $\hbrho$ satisfying \eref{eq:hamilton_coords} and \eref{eq:hamilton_momenta}. Since $\chi$ is positive-definite, the Lagrangian is zero only at $\hbrho = 0$ and otherwise positive. Since trajectories with nonzero energy always have $\hbrho \neq 0$ (even in the limit $t \to \pm \infty$), the action \eref{eq:action_lagrangian} associated with these trajectories keeps increasing in time without any upper bound. On the other hand, trajectories with zero energy have limit sets satisfying $\hbrho = 0$. Therefore their action saturates to a finite value in the limit $t_\rmi \to -\infty$ or $t_\rmf \to +\infty$.

\end{enumerate}

\subsubsection{Evaluation of the LDF} \label{sssec:evaluation}

With these observations, a method for evaluating the LDF $\phi$ can be derived. For any time interval bounded by $t_\rmi$ and $t_\rmf$, the steady-state distribution $P_{\rm s}(\brho)$ satisfies
\begin{equation} \label{eq:steady1}
P_{\rm s}(\brho) = \int \rmd \brho_\rmi \, P(\brho,t_\rmf|\brho_\rmi,t_\rmi) P_{\rm s}(\brho_\rmi).
\end{equation}
Using \eref{eq:ld_principle} and \eref{eq:propagator} with the saddle-point approximation, this equation can be rewritten as
\begin{equation} \label{eq:ldf_steady1}
\phi(\brho) = \min_{\brho_\rmi} \left[\phi(\brho_\rmi) + S(\brho,t_\rmf;\brho_\rmi,t_\rmi)\right].
\end{equation}
As is clear from \eref{eq:lagrangian_momenta}, we always have $S(\brho,t_\rmf;\brho_\rmi,t_\rmi) \ge 0$. Thus \eref{eq:ldf_steady1} is consistent with our previous statement that the attractors of the noiseless dynamics correspond to local minima of $\phi$. Before proceeding further, we note that as far as quantities derived from the action are concerned, any pair of configurations on the {\it same limit set} can be regarded as completely equivalent to each other. This is because such configurations are mutually accessible by trajectories with $\hbrho = 0$, which do not cost any action according to \eref{eq:action} and \eref{eq:lagrangian_momenta}. Thus, whenever a configuration belonging to a limit set $\Gamma_\alpha$ appears in the argument of an action-related quantity, we can replace the configuration with $\Gamma_\alpha$ without any ambiguities.

We choose $t_\rmi = -\infty$ and $t_\rmf = 0$ for \eref{eq:ldf_steady1}, which conveniently limits the relevant trajectories to those with zero energy; trajectories of nonzero energy, as previously observed, cost an infinite positive action. Therefore, only trajectories of zero energy, whose initial configuration $\brho_\rmi$ approaches the limit sets of the noiseless dynamics, can possibly minimize the action. We introduce a notation $\alpha \in {\cal N}^- (\brho)$, which indicates that the limit set $\Gamma_\alpha$ is reached from $\brho$ in the limit $t_\rmi \to -\infty$. Then we can rewrite \eref{eq:ldf_steady1} as
\begin{equation} \label{eq:ldf1}
\phi(\brho) = \min_{\alpha \in {\cal N}^- (\brho)} \left[\phi(\Gamma_\alpha) + S(\brho,0;\Gamma_\alpha,-\infty)\right],
\end{equation}
which still leaves the values of $\phi$ on the limit sets undetermined.

In order to fix the values of the LDF $\phi$ on the limit sets, we use the fact that in the steady state
\begin{equation} \label{eq:steady2}
\int_{\brho' \neq \brho} \rmd \brho' \, P(\brho',t_\rmf|\brho,t_\rmi) P_{\rm s}(\brho) = \int_{\brho' \neq \brho} \rmd \brho' \, P(\brho,t_\rmf|\brho',t_\rmi) P_{\rm s}(\brho').
\end{equation}
In other words, the probability flux out of $\brho$ must cancel the probability flux into it. This time we use the prescription $t_\rmi = -\infty$ and $t_\rmf = +\infty$, which ensures that $\brho$ and $\brho'$ are a pair of neighboring limit sets, as discussed in point (ii) of \ref{sssec:structure}. Thus, we obtain
\begin{equation} \label{eq:ldf2}
\fl \phi(\Gamma_\alpha) + \min_{\beta \in {\cal N}^+(\alpha)} [S(\Gamma_{\beta},\infty;\Gamma_\alpha,-\infty)] = \min_{\beta \in {\cal N}^-(\alpha)}\left[\phi(\Gamma_{\beta}) + S(\Gamma_{\alpha},\infty;\Gamma_{\beta},-\infty)\right],
\end{equation}
which provides self-consistent relations for the values of $\phi$ on the limit sets (see also Chapter 6 of \cite{Freidlin2012} for a more rigorous discussion). Combining \eref{eq:ldf1} and \eref{eq:ldf2}, we can determine $\phi$ for any point in the configuration space up to an undetermined global additive constant. The constant can be set by choosing the global additive constant so that $\phi(\Gamma^*) = 0$ at the global minimum of $\phi$, where $\Gamma^*$ corresponds to the most probable state of system.

\subsection{Alternative approach: Fokker--Planck equations}

It is instructive to see that the same results can also be derived by noting that the Langevin equation (\ref{eq:langevin}) is equivalent to the Fokker--Planck equation
\begin{equation}
\frac{\partial}{\partial t} P(\brho,t) =
\left[-\frac{\partial}{\partial \brho} \cdot \bi{F}(\brho) + \frac{1}{2N} \sum_{\mu,\nu} \frac{\partial^2}{\partial \rho_\mu \partial \rho_\nu} \chi_{\mu\nu}(\brho) \right] P(\brho,t).
\end{equation}
When the system reaches the steady state, we may use the large deviation principle given by \eref{eq:ld_principle} in place of $P(\brho,t)$. Leaving only the leading-order terms, we obtain
\begin{equation} \label{eq:ldf_pde}
0 = \bi{F}(\brho) \cdot \frac{\partial \phi}{\partial \brho} + \frac{1}{2}\frac{\partial \phi}{\partial \brho} \cdot \chi(\brho) \frac{\partial \phi}{\partial \brho}.
\end{equation}
A comparison with the Hamiltonian defined by \eref{eq:hamiltonian} shows that this equation can be rewritten as
\begin{equation} \label{eq:hamilton_jacobi}
H\left(\brho,\frac{\partial \phi}{\partial \brho} \right) = 0,
\end{equation}
which can be regarded as a Hamilton--Jacobi equation for zero energy~\cite{Arnold1989}. This interpretation allows us to apply the standard formalism of analytical mechanics, which assigns the meaning of an action (possibly modified by an additive constant) to $\phi$ and
introduces momenta by the relation
\begin{equation} \label{eq:momenta_action}
\hbrho = \frac{\partial \phi}{\partial \brho}.
\end{equation}
This relation is consistent with \eref{eq:ldf1} if $\hbrho$ is interpreted as the final momentum of a trajectory that starts from a limit set and dominates $\phi$. Proceeding from this interpretation, we can derive the Hamiltonian structure previously discussed.

\subsection{Systems with detailed balance} \label{ssec:db}

If the system is in equilibrium and thus satisfies detailed balance (DB), its LDF $\phi$ can be obtained in a simpler way. For the sake of simplicity, we assume that all system variables are even (i.e. do not change sign) under time reversal. Given this assumption, DB requires that the flux of probability between any pair of (possibly equal) configurations $\brho$ and $\brho'$ satisfies~\cite{Haken1975}
\begin{equation} \label{eq:db}
P(\brho',t_\rmf|\brho,t_\rmi) \, P_{\rm s}(\brho) = P(\brho,t_\rmf|\brho',t_\rmi) \, P_{\rm s}(\brho').
\end{equation}
We consider the case of $t_\rmf = t_\rmi +\rmd t$ with $\rmd t \ll 1$, so that the system evolves only for an infinitesimal time interval. Due to the large deviation principle \eref{eq:ld_principle}, the steady-state probability of $\brho'$ can be written as
\begin{equation} \label{eq:ldf_fin}
P_{\rm s}(\brho') \sim \exp [-N\phi(\brho')] \sim \exp \left\{-N\left[ \phi(\brho) + \frac{\partial \phi}{\partial \brho} \cdot \dot{\brho} \, \rmd t \right]\right\}.
\end{equation}
Meanwhile, the ``forward'' propagator from $\brho$ to $\brho'$ in the LHS of \eref{eq:db} is given by \eref{eq:propagator}, with the action evaluated by the constraint of zero energy. Thus, we obtain
\begin{equation} \label{eq:forward}
P(\brho',t_\rmf|\brho,t_\rmi) \sim \exp \left\{-N\dot{\brho} \cdot \hbrho \, \rmd t \right\},
\end{equation}
where $\dot{\brho} = (\brho' - \brho)/\rmd t$ and $\hbrho$ satisfies the equation of motion \eref{eq:hamilton_coords}. The ``backward'' propagator from $\brho'$ to $\brho$ in the RHS of \eref{eq:db} can be obtained in a similar way, after applying appropriate time reversal operations $\dot{\brho} \to -\dot{\brho}$ and $\hbrho \to \hbrho^{\rm TR}$ so that the time-reversed equation of motion
\begin{equation} \label{eq:hamilton_coords_tr}
-\dot{\brho} = \bi{F}(\brho) + \chi(\brho)\hbrho^{\rm TR}
\end{equation}
is satisfied. Then, in a manner similar to \eref{eq:forward}, we have
\begin{equation} \label{eq:backward}
P(\brho,t_\rmf|\brho',t_\rmi) \sim \exp \left\{N\dot{\brho} \cdot \hbrho^{\rm TR} \, \rmd t \right\}.
\end{equation}
Using \eref{eq:ldf_fin}, \eref{eq:forward}, and \eref{eq:backward} in \eref{eq:db}, we obtain
\begin{equation} \label{eq:ldf_db0}
\frac{\partial \phi}{\partial \brho} = \hbrho + \hbrho^{\rm TR}.
\end{equation}
This expression may appear to contradict \eref{eq:momenta_action}, but \eref{eq:ldf_db0} is about trajectories between any pair of nearby configurations $\brho$ and $\brho'$, while in \eref{eq:momenta_action} $\hbrho$ is the final momentum of a trajectory from a limit set. From \eref{eq:hamilton_coords} and \eref{eq:hamilton_coords_tr}, we rewrite \eref{eq:ldf_db0} as
\begin{equation} \label{eq:ldf_db}
\frac{\partial \phi}{\partial \brho} = - 2 \chi^{-1} (\brho) \bi{F}(\brho).
\end{equation}
The LDF $\phi$ can be determined simply by integrating the RHS of this equation. We note that \eref{eq:ldf_db} is a sufficient condition for the Hamilton--Jacobi equation \eref{eq:hamilton_jacobi}, which confirms that \eref{eq:ldf_db} is indeed consistent with the Hamiltonian structure discussed above. Moreover, since we have already assumed that $\chi$ and $\bi{F}$ are smooth and that $\chi$ is invertible, \eref{eq:ldf_db} implies that $\phi$ is also a smooth function.

The above results can be easily generalized to cases involving odd-parity variables. See \cite{Graham1971a, Graham1971b, Risken1972} for general approaches based on the Fokker--Planck equations.

\section{Singularities in finite-dimensional systems} \label{sec:fin_dim}

If a system is out of equilibrium, the DB condition \eref{eq:ldf_db} is not satisfied. Then, even if ${\bf F}$ and $G$ are smooth, the LDF of the system is not necessarily smooth in the weak-noise limit; it is well known that the function may have singularities \cite{Graham1971a, Graham1971b, Risken1972}. For finite-dimensional systems, the origins of such singularities are well understood from the Hamiltonian structure discussed in \sref{sec:formalism}. Broadly, there are two major scenarios which lead to singular LDFs. The first originates from the existence of multiple limit sets in the system. In this case there are generically several locally minimizing trajectories, each originating from a different limit set, which lead to the same final configuration with the {\it globally minimizing} one dictating the LDF (see below). As the configuration is changed, the locally minimizing trajectories change smoothly. However, there are specific points where the globally minimizing trajectory changes by originating from a different limit set. This, much like a usual first order phase transition, leads to a singular LDF. In a second scenario, locally minimizing trajectories {\it originating from the same limit set} compete and lead to a singular behavior. In this case a singular LDF appears as a bifurcation, and for reasons that will become clear later we refer to these as Landau-like singularities.

In the following we discuss how each of these scenarios gives rise to singularities in finite-dimensional systems. We also touch upon other types of singularities that arise in more exceptional cases.

\subsection{Singularities due to multiple limit sets}
\label{ssec:singular_limit_sets}

As expressed in \eref{eq:ldf1}, the LDF $\phi$ is evaluated by choosing the limit set that minimizes its value. As discussed above, in the presence of multiple limit sets, a singularity of $\phi$ can occur at points where the minimizing limit set changes abruptly as $\brho$ is continuously varied.

\begin{figure}
\begin{center}
\includegraphics[height=1.0488in,width=1.7113in]{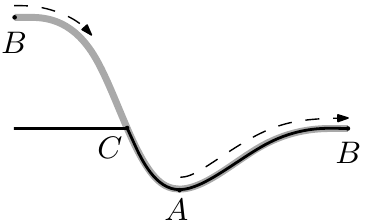}
\caption{\label{fig:multiple_limit_sets} A simple model with a singularity of the LDF due to multiple limit sets. The gray curve represents the effective potential $U$, and the black curve shows the LDF $\phi$ (see the main text for definitions). For the flat portion of $\phi$, the trajectories to the right (marked by dashed arrows) are less costly than those to the left in terms of action.}
\end{center}
\end{figure}

We present a simple example of such singularities, which is adapted from \cite{Graham1986}. Consider an overdamped particle on a one-dimensional ring, whose position $\rho$ evolves according to the Langevin equation
\begin{equation} \label{eq:ex_limit_set_langevin}
\dot{\rho} = f - V'(\rho) + \eta(t).
\end{equation}
Here $f > 0$ is a constant force, $V$ a smooth periodic potential, and $\eta$ a Gaussian white noise satisfying $\langle \eta(t) \eta(t') \rangle = 2N^{-1}\delta(t-t')$. See \fref{fig:multiple_limit_sets} for an illustration of this system. We assume that the effective potential $U(\rho) \equiv -f\rho + V(\rho)$, represented by the gray curve in \fref{fig:multiple_limit_sets}, satisfies $U'(\rho) = 0$ only on two points $A$ and $B$ with $U''(A) < 0$ and $U''(B) > 0$. Thus, A (B) is the attractor (repeller) of the noiseless dynamics corresponding to \eref{eq:ex_limit_set_langevin}.

In the large $N$ limit, the system is related to a mechanical system characterized by the Hamiltonian
\begin{equation}
H(\rho,\hat{\rho}) = -\hat{\rho}\,U'(\rho) + \hat{\rho}^2,
\end{equation}
which indeed has the form of \eref{eq:hamiltonian}. As discussed in \ref{sssec:evaluation}, only trajectories with $H(\rho,\hat{\rho}) = 0$ contribute to the LDF. Thus, we can limit our attention to two types of trajectories given by $\hat{\rho} = 0$ and $\hat{\rho} = U'(\rho)$. ``Downhill'' trajectories from $B$ correspond to the former and do not cost any action, satisfying $S(\rho_\rmf,0|B,-\infty) = 0$. On the other hand, ``uphill'' trajectories from $A$ belong to the latter type, whose action is given by
\begin{equation}
S(\rho_\rmf,0|A,-\infty) = \int_{-\infty}^0 \rmd t \, \hat{\rho} \dot{\rho} = \int_{A}^{\rho_\rmf} \rmd \rho \, U'(\rho) > 0,
\end{equation}
where we used the zero-energy constraint for the first equality and used the substitution $\dot{\rho} \, \rmd t = \rmd \rho$ for the second. Using \eref{eq:ldf2}, we can fix the difference between $\phi(A)$ and $\phi(B)$ by
\begin{equation}
\fl \phi(B) = \phi(A) + \min \left[ \int_{ACB} \rmd \rho \, U'(\rho) , \int_{AB} \rmd \rho \, U'(\rho) \right] = \phi(A) + \int_{AB} \rmd \rho \, U'(\rho),
\end{equation}
where $ACB$ ($AB$) represents the trajectory from $A$ to $B$ that is headed to the left (right). As is evident from \fref{fig:multiple_limit_sets}, the trajectory $AB$ costs less action and dominates the LDF. Since $A$ is the most probable position, we fix the zero point of the LDF by setting $\phi(A) = 0$. Finally, the LDF at any position $\rho_\rmf$ is determined from \eref{eq:ldf1} as
\begin{equation} 
\phi(\rho_\rmf) = \min \left[\int_{AB} \rmd \rho \, U'(\rho), \int_{A}^{\rho_\rmf} \rmd \rho \, U'(\rho) \right].
\end{equation}
Note that each option corresponds to a different limit set: the first to $B$ and the second to $A$. As shown in \fref{fig:multiple_limit_sets}, there exists a point $C$ at which the minimizing limit set changes, so that the LDF has the form
\begin{equation}
\phi(\rho_\rmf) = \cases{
\int_{AB} \rmd \rho \, U'(\rho) &if $\rho_\rmf$ is between $B$ and $C$\\
\int_{A}^{\rho_\rmf} \rmd \rho \, U'(\rho) &otherwise}
\end{equation}
as represented by the black curve in \fref{fig:multiple_limit_sets}. Then the first derivative of $\phi$ has a discontinuity at $C$. This singularity reflects the fact that the barrier on the left is larger than the barrier on the right, which makes it easier to reach the interval between $B$ and $C$ by trajectories going to the right, which are marked by representative dashed arrows in \fref{fig:multiple_limit_sets}. It is easy to verify that this singular behavior is lost for $f = 0$, i.e. when the system is in equilibrium.

The same type of singularities have been found in various model systems. An exactly solvable model studied by \cite{Hu1988,Hu1989} has a separable angular degree of freedom, whose singular behavior is essentially the same as that of our example. Similar examples are also discussed in \cite{Ge2012}. An underdamped counterpart of our example, which models a dc-biased Josephson junctions and charge density waves, was also found to have a line of singularities separating a stable fixed point from a saddle point or a stable limit cycle~\cite{Graham1986,BenJacob1982}. In addition to these periodic systems, a non-equilibrium system with symmetric double wells~\cite{Maier1993a, Maier1993b, Maier1996a, Maier2000} also exhibits such singularities along the boundary between the two basins of attraction.

Singularities due to multiple limit sets are generally of a first-order nature; the first derivative of the LDF is singular. In what follows we discuss other kinds of singularities, which turn out to have a structure analogous to phase transitions described by a Landau theory.

\subsection{Landau-like singularities}
\label{ssec:singular_landau}

\subsubsection{Bifurcation of locally minimizing trajectories} \label{sssec:qs}

Even if the system has a single limit set, the LDF may still possess singularities. In principle, a single initial limit set (even in the case when it is a fixed point) can give rise to multiple locally minimizing trajectories that reach a final configuration $\rho_\rmf$. When such cases arise, the least action principle \eref{eq:action} picks the trajectory with the minimal action as the globally minimizing trajectory that determines the LDF. As $\rho_\rmf$ is continuously varied, the globally minimizing trajectory may switch abruptly from one to another, creating a boundary on which multiple globally minimizing trajectories coexist. The LDF becomes non-differentiable on such boundaries. As we show below, this can only occur when $\rho_\rmf$ is far enough from the initial limit set, so that there is a bifurcation of locally minimizing trajectories at some ``threshold'' of $\rho_\rmf$.

\begin{figure}
\begin{center}
\includegraphics[width=0.45\textwidth]{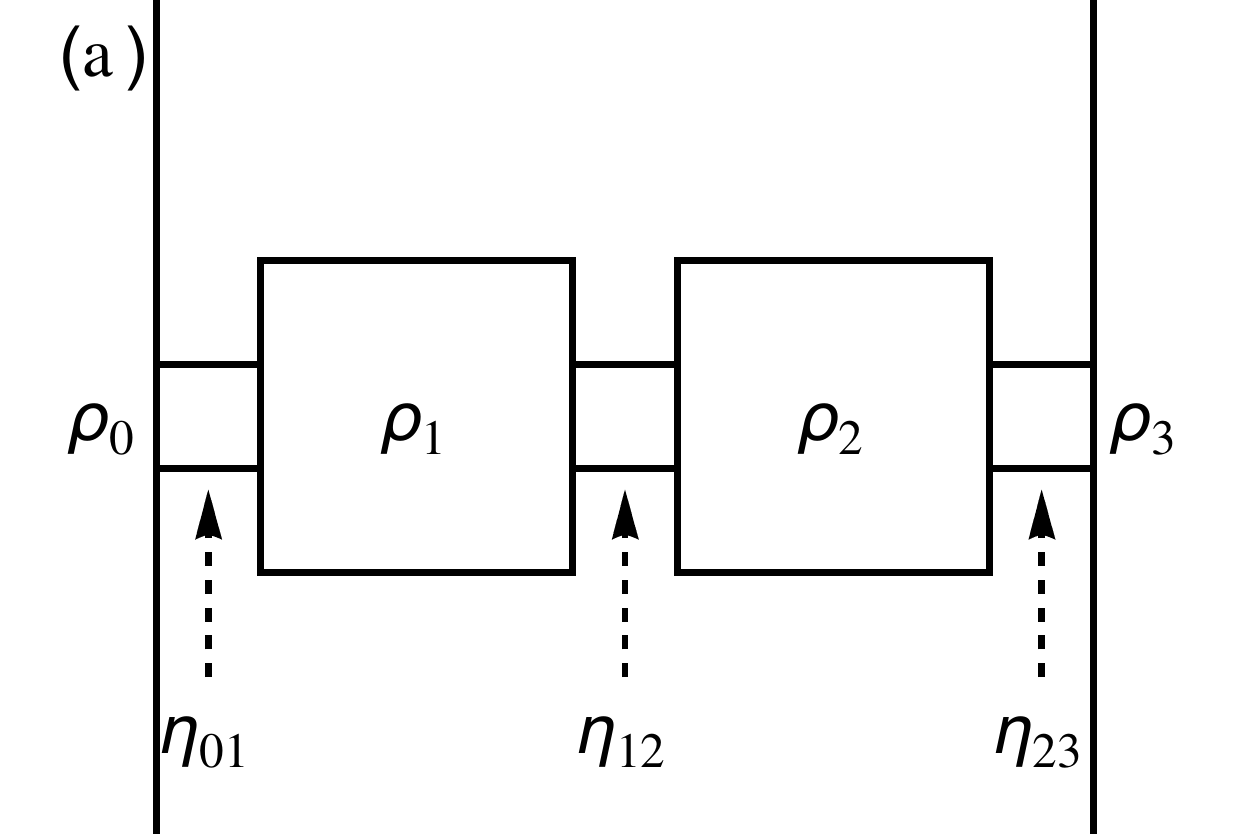}\quad
\includegraphics[width=0.45\textwidth]{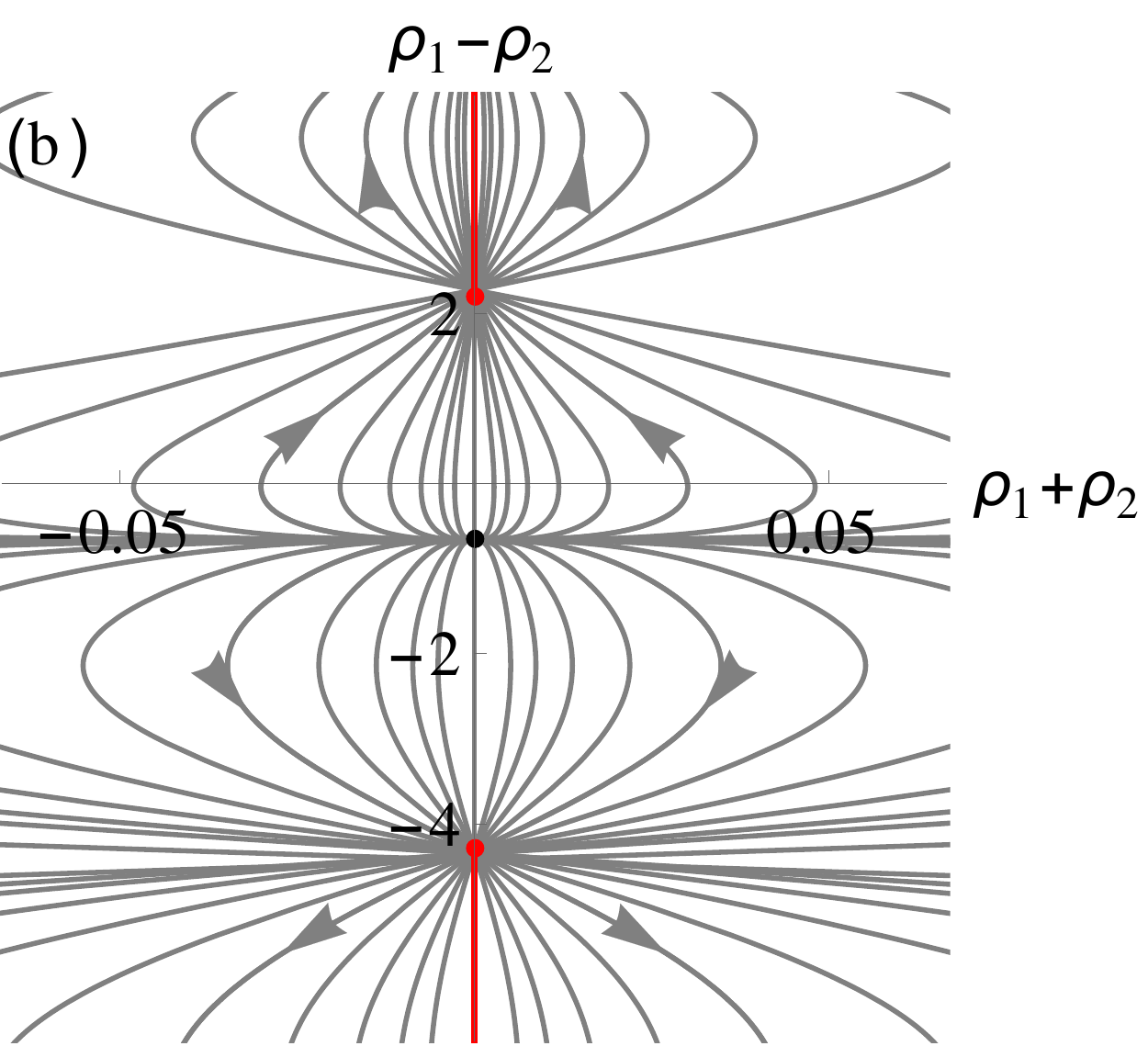}
\end{center}
\caption{\label{fig:qs} (a) A two-component boundary-driven system described by \eref{eq:qs_langevin}. (b) The extremizing trajectories (gray curves) originating from the stable fixed point (black point), which are obtained from the Hamiltonian \eref{eq:qs_hamiltonian} with $\bar{\rho} = 1$. A pair of red points (lines) indicate the bifurcation points (switching lines).}
\end{figure}

To provide intuition into how such bifurcation can occur, we present a simple example: a discretized version of a transport model between two reservoirs~\cite{Bertini2005b} which is known to exhibit this type of singularities in driven diffusive systems~\cite{Bunin2013a}. Consider a two-component system described by the density variables $\rho_1$ and $\rho_2$, which exchange current with each other as well as the reservoirs of densities that are respectively fixed at $\rho_0$ and $\rho_3$ [see \fref{fig:qs}(a)]. We assume that the current between neighboring components fluctuates so that $\rho_1$ and $\rho_2$ evolve by the Langevin equations
\begin{eqnarray} \label{eq:qs_langevin}
\dot{\rho}_1 &= \rho_0 - 2\rho_1 + \rho_2 + \sqrt{\sigma_{01}(\rho_0,\rho_1)}\,\eta_{01} -  \sqrt{\sigma_{12}(\rho_1,\rho_2)}\,\eta_{12}, \nonumber \\
\dot{\rho}_2 &= \rho_1  - 2\rho_2 +\rho_3 + \sqrt{\sigma_{12}(\rho_1,\rho_2)}\,\eta_{12} - \sqrt{\sigma_{23}(\rho_2,\rho_3)}\,\eta_{23},
\end{eqnarray}
where the noise amplitudes satisfy
\begin{equation}
\sigma_{\mu\nu}(\rho_\mu,\rho_\nu) = 1 + (\rho_\mu + \rho_\nu)^2, \quad \langle \eta_\mu (t) \eta_{\nu} (t')\rangle = N^{-1} \delta_{\mu\nu} \delta(t-t')
\end{equation}
for any two indices $\mu$ and $\nu$. In the absence of noise, the system relaxes to the stable fixed point at $\brho_* \equiv (\frac{2\rho_0+\rho_3}{3},\frac{\rho_0+2\rho_3}{3})$, which is the only limit set of the dynamics. It is easy to verify that as long as $\rho_0 \neq \rho_3$ the system carries a current and is out of equilibrium. For simplicity, we consider the special case when the reservoirs satisfy $\rho_3 = -\rho_0 = \bar{\rho}$, which drives the system out of equilibrium but still maintains some symmetry. Then, the Hamiltonian describing the large $N$ limit of this system is obtained from \eref{eq:hamiltonian}:
\begin{eqnarray}
\fl H(\brho,\hbrho) = \hrho_1 \, (\rho_2-\bar{\rho} - 2\rho_1) + \hrho_2 \, (\rho_1 + \bar{\rho} - 2\rho_2) \nonumber \\
\fl \qquad + \frac{1}{2} \, \hrho_1^2 \, \left[1+(\rho_1-\bar{\rho})^2\right] + \frac{1}{2} \, (\hrho_1-\hrho_2)^2 \, \left[1+(\rho_1+\rho_2)^2\right] + \frac{1}{2} \, \hrho_2^2\,\left[1+(\rho_2+\bar{\rho})^2\right] \label{eq:qs_hamiltonian}.
\end{eqnarray}
The corresponding equations of motion can be obtained from \eref{eq:hamilton_coords} and \eref{eq:hamilton_momenta}, which can be solved numerically and together with the least action principle \eref{eq:action} give rise to the extremizing trajectories shown in \fref{fig:qs}(b). The trajectories show that, if the final configuration $\brho_\rmf$ is on the line $\rho_1 + \rho_2 = 0$ and sufficiently far away from $\brho_*$, there are two histories deviating from $\rho_1 + \rho_2 = 0$. By evaluating the action accumulated during each history, one can see that these histories indicate a pair of globally minimizing trajectories from $\brho_*$ to $\brho_\rmf$. The collection of $\brho_\rmf$ with multiple globally minimizing trajectories form two semi-infinite lines, as marked by the red lines in \fref{fig:qs}(b). These lines, often called {\it switching lines}, indicate the boundaries on which the globally minimizing trajectory switches abruptly. Meanwhile, if $\brho_\rmf$ satisfies $\rho_1 + \rho_2 = 0$ but is close enough to $\brho_*$, there is only a single globally minimizing trajectory that stays on the straight line $\rho_1 + \rho_2 = 0$. This can be understood by noting that close to $\brho_*$ the equations of motion can be linearized. The solutions of the linearized equations can never lead to coexisting trajectories.

The origin of the bifurcation can be intuitively understood from the form of the Langevin equation \eref{eq:qs_langevin}. The excitation of the system along the line $\rho_1 + \rho_2 = 0$ requires that $\rho_1$ and $\rho_2$ evolve in opposite directions, which is efficiently driven by the noise component $\eta_{12}$ thanks to its opposite signs in \eref{eq:qs_langevin}. Note that the amplitude $\sigma_{12}$ of $\eta_{12}$ is minimized along the line $\rho_1 + \rho_2 = 0$. Thus, introducing a small deviation from $\rho_1 + \rho_2 = 0$ along the trajectory increases the driving provided by $\eta_{12}$, which for sufficiently far away $\brho_\rmf$ may reduce the action despite the increased length of the trajectory to the final configuration. Deviations in both $\rho_1 + \rho_2 > 0$ and $\rho_1 + \rho_2 < 0$ directions have a symmetric ``boost'' of the drive, so the globally minimizing trajectories have a reflection symmetry with respect to the $\rho_1 + \rho_2 = 0$ line.

This example indicates that the end point of a switching line indicates the point at which different locally minimizing trajectories start to bifurcate. The structure of such bifurcations can be described by an analogue of a Landau theory, which we now discuss in more detail.

\subsubsection{Formulation of the Landau theory} \label{sssec:landau}
\begin{figure}
\begin{center}
\includegraphics[width=0.5\textwidth]{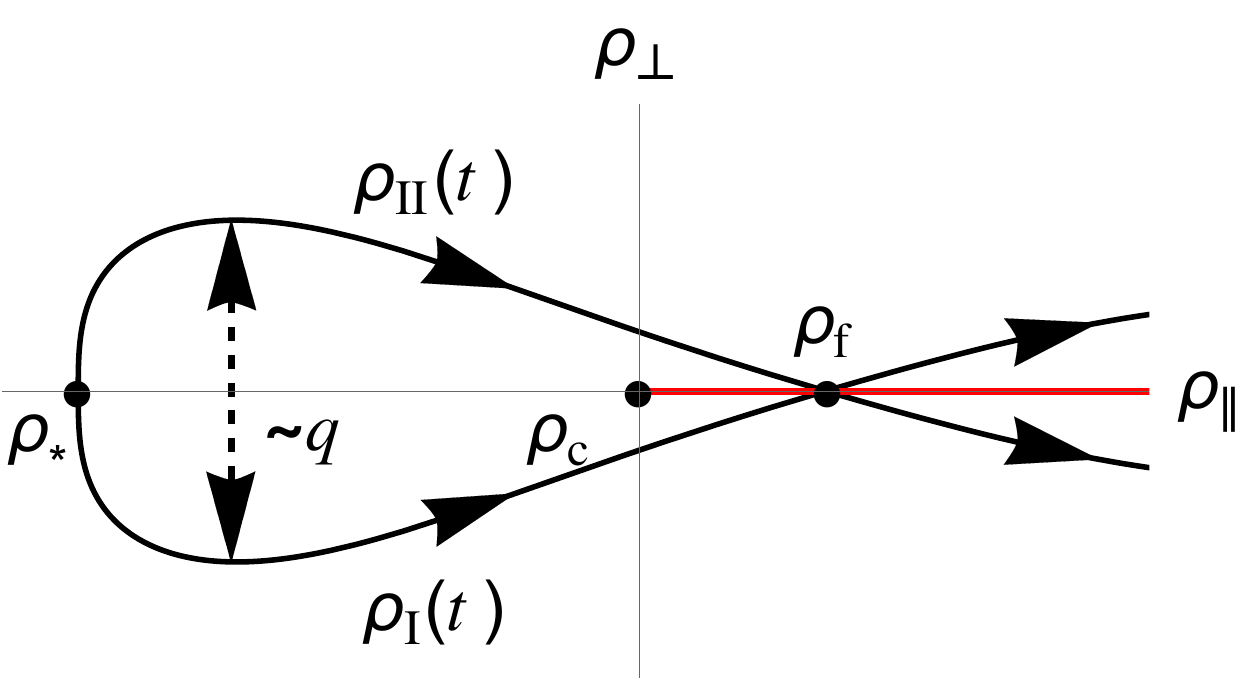}
\caption{\label{fig:landau_parameters} Geometric interpretations of the parameters appearing in the Landau free energy \eref{eq:landau}. The red solid line represents the switching line.}
\end{center}
\end{figure}

We present a phenomenological formulation of the Landau theory for trajectory bifurcations. We assume that the system has a unique stable fixed point $\brho_*$ and that locally minimizing trajectories from $\brho_*$ start to bifurcate when the final configuration is at $\brho_{\rm c}$. It is convenient to introduce new coordinates $(\rho_\parallel,\rho_\perp)$ for the configuration space, so that the origin is located at $\brho_{\rm c}$, the positive $\rho_\parallel$-axis coincides with the switching line, and the $\rho_\perp$-axis is perpendicular to the switching line. If there are multiple possible axes perpendicular to the $\rho_\parallel$-axis, the $\rho_\perp$-axis is chosen so that it represents the direction in which trajectories move away from each other as they bifurcate. In the example of \eref{eq:qs_langevin}, the $\rho_\parallel$-axis runs parallel to the lines of constant $\rho_1 + \rho_2$, and the $\rho_\perp$-axis is parallel with the lines of constant $\rho_1 - \rho_2$.

Let $\brho_{\rm I}(t)$ and $\brho_{\rm II}(t)$ denote (possibly identical) locally minimizing trajectories reaching the final configuration $(\rho_\parallel,\rho_\perp)$. We define
\begin{equation} \label{eq:traj_def}
\brho_{\rm avg}(t) \equiv \frac{\brho_{\rm I}(t) + \brho_{\rm II}(t)}{2},\quad \delta\brho(t) \equiv \frac{\brho_{\rm I}(t) - \brho_{\rm II}(t)}{2},
\quad \bi{u}(t) \equiv \frac{\delta \brho(t)}{\|\delta \brho(t)\|},
\end{equation}
where $\|\delta \brho(t)\|^2 \equiv \int_{-\infty}^0 \rmd t \, \delta \brho(t) \cdot \delta \brho(t)$. See \fref{fig:landau_parameters} for an illustration of these trajectories. With these definitions, we formulate a Landau free energy ${\cal L}(q;\rho_\parallel,\rho_\perp)$, so that the order parameter $q$ minimizing $\cal L$ captures the behavior of $\|\delta \brho(t)\|$, which becomes zero before the bifurcation ($\rho_\parallel < 0$) and nonzero after the bifurcation ($\rho_\parallel > 0$). More specifically, we consider $q$ that satisfies $q \sim \|\delta \brho(t)\|$ close to the bifurcation (see \fref{fig:landau_parameters}). The appropriate form of $\cal L$ is
\begin{equation} \label{eq:landau}
{\cal L}(q;\rho_\parallel,\rho_\perp) = c_4 q^4 - c_2 \rho_\parallel q^2 + \rho_\perp q,
\end{equation}
where $c_2$ and $c_4$ are positive coefficients. Using the condition that $\cal L$ must be minimized, we obtain a ``phase diagram'' in the $(\rho_\parallel,\rho_\perp)$-space, as shown in \fref{fig:lagrangian_manifold}(b). For $\rho_\parallel > 0$ and $\rho_\perp^2 < (8 c_2^3 \rho_\parallel^3)/(27c_4)$, there exist two different local minima of $\cal L$, which indicates the presence of multiple locally minimizing trajectories in the region. Interpreting these trajectories as ``metastable states'', the boundaries of the region given by $\rho_\perp^2 = (8 c_2^3 \rho_\parallel^3)/(27c_4)$ can be regarded as {\it spinodal lines}, which are indicated by the thick black lines in \fref{fig:lagrangian_manifold}(b). Within the region bounded by the spinodal lines, both locally minimizing trajectories have the same value of $\cal L$ on the switching line $\rho_\perp = 0$ (the thick red line of \fref{fig:lagrangian_manifold}), which indicates that they are equally dominant. The abrupt change of the minimizing $q$ as the sign of $\rho_\perp$ flips across this line means that $\cal L$ has a discontinuity in its first derivative across the line [see \fref{fig:lagrangian_manifold}(c)], so the switching line becomes an analogue of the {\it discontinuous transition line}. Finally, the common end point of all spinodal lines and the discontinuous transition line at the origin $\rho_\parallel = \rho_\perp = 0$ is naturally interpreted as a {\it critical point}. This predicts $q \sim \rho_\parallel^{1/2}$ for the separation between bifurcating trajectories for small $\rho_\parallel$.

We have explained how singularities associated with trajectory bifurcations can be described by a Landau theory on a phenomenological level. We note that $\cal L$ can be derived more systematically by approximating the path integrals of \sref{sec:formalism} close to $\brho_{\rm c}$ and integrating them over variables which are not directly related to the trajectory bifurcation. See Chapter 15 of \cite{Schulman1981} for details. Such derivations show that the steady-state distribution can be written in the form
\begin{equation} \label{eq:steady_pdf_q}
P_{\rm s} (\brho_\rmf) \sim \exp \left[-N\phi(\brho_{\rm c})\right] \int \rmd q \, \exp \left[ -N {\cal L}(q;\rho_\parallel,\rho_\perp) \right].
\end{equation}
Applying the saddle-point method, the value of $q$ must be such that $\cal L$ is minimized. From \eref{eq:ld_principle}, it is clear that the minimum of $\cal L$ is an additive term of the LDF $\phi$. Therefore any singularity arising from the minimization of $\cal L$ captures the singular behavior of $\phi$.

Finally, we note that \eref{eq:landau} is not the only possible form of $\cal L$. Depending on the system, the trajectory bifurcation may involve more than two globally minimizing trajectories, in which case it is better described by the Landau theory of multicritical points. Catastrophe theory has developed a series of standard forms of $\cal L$, among which \eref{eq:landau} is a special case usually referred to as a {\it cusp catastrophe}~\cite{Arnold1992}. We will see later that there are indeed models whose Landau-like singularities are described by the other forms of $\cal L$.

\subsubsection{Connections with Lagrangian manifolds} \label{sssec:landau_lm}

Landau-like singularities are related to a nontrivial geometry of a manifold in the phase space which is formed by zero-energy trajectories. This is referred to as a {\it Lagrangian manifold} (LM). More specifically, the relevant LM is formed by trajectories originating from an attractor, called an {\it unstable} LM. The structure of the unstable LM provides a deeper understanding of the patterns of extremizing trajectories close to a Landau-like singularity.

A simple example of such LM can be obtained from \eref{eq:steady_pdf_q}. We observe that $\cal L$ plays the role of an action, whose minimum value contributes to the LDF $\phi$. The momentum $\hrho_\perp$ conjugate to the coordinate $\rho_\perp$ is
\begin{equation}
\hrho_\perp = \frac{\partial}{\partial \rho_\perp} {\cal L}(q;\rho_\parallel,\rho_\perp) = q.
\end{equation}
This allows us to interpret the order parameter as a momentum component associated with a direction perpendicular to the switching line. Note that the other momentum component,
\begin{equation}
\hrho_\parallel = \frac{\partial \phi}{\partial \rho_\parallel} = -c_2q^2,
\end{equation}
is equal for any pair of two globally minimizing trajectories meeting on the switching line. Thus, $\hrho_\perp$ is the only momentum component that differs between coexistent globally minimizing trajectories at the final time. From the condition $\partial {\cal L}/\partial q = 0$ for the minimization of $\cal L$, we obtain
\begin{equation} \label{eq:lagrangian_manifold}
\rho_\perp = 2c_2 \rho_\parallel \hrho_\perp -  4c_4 \hrho_\perp^3,
\end{equation}
which gives the structure of the LM close to $\brho_{\rm c}$.

\begin{figure}
\begin{center}
\includegraphics[width=0.7\textwidth]{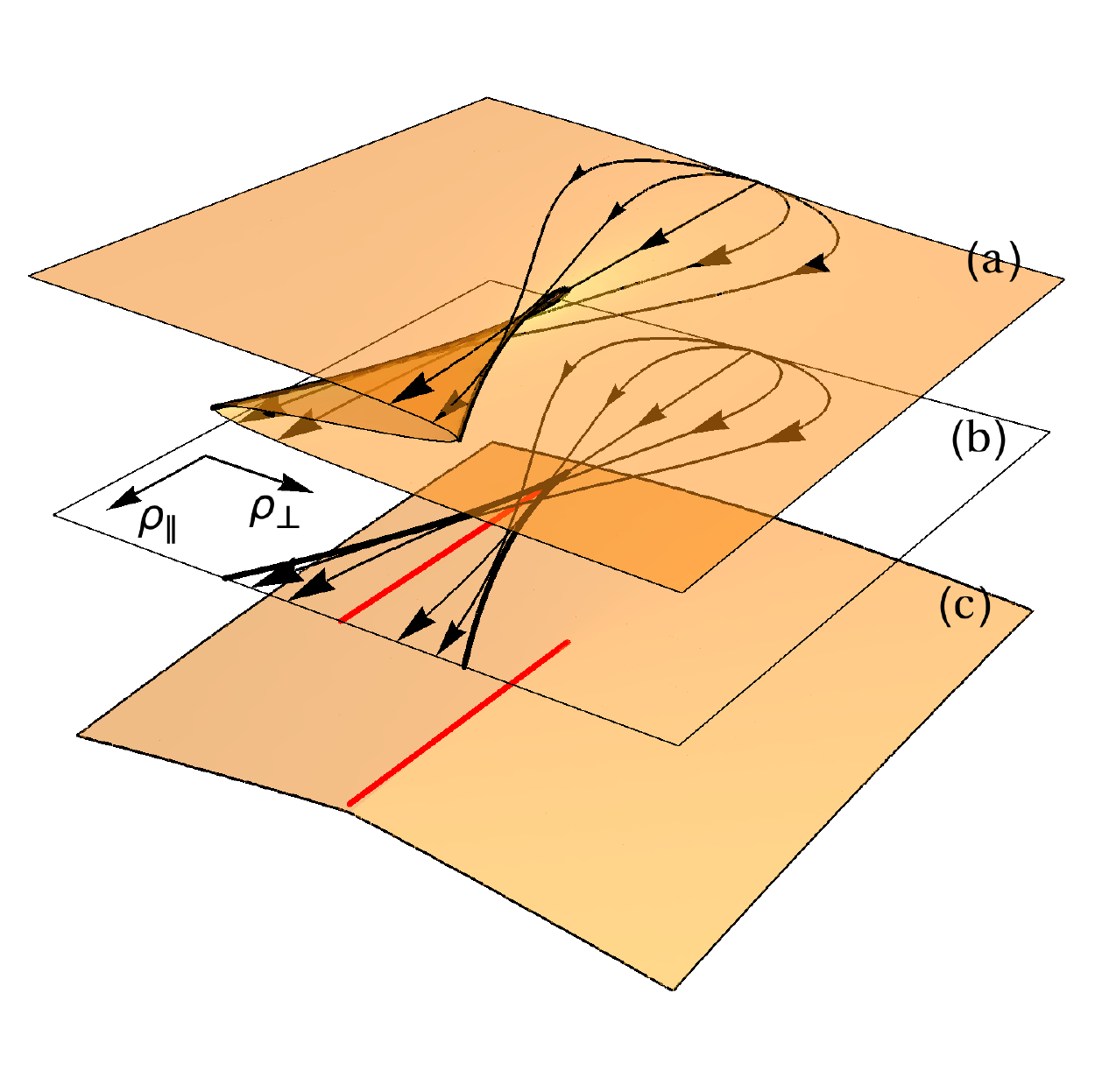}
\end{center}
\caption{\label{fig:lagrangian_manifold} (a) The typical structure of the LM close to a cusp point. The thin black lines with arrows indicate extremizing trajectories, and the thick black lines correspond to fold lines. (b) A projection of the LM on the configuration space yields a ``phase diagram'' of coexisting trajectories, in which the thick black lines are spinodal lines and the thick red line is the discontinuous transition line. (c) The corresponding LDF near the same cusp point has a non-differentiable line (thick red line).}
\end{figure}

Figure \ref{fig:lagrangian_manifold}(a) shows the schematic shape of the LM satisfying \eref{eq:lagrangian_manifold}. The LM has multiple sheets in the region satisfying $\rho_\parallel > 0$ and $\rho_\perp^2 \le (8 c_2^3 \rho_\parallel^3)/(27c_4)$; in the rest of the configuration space, the LM is a single sheet. The boundaries $\rho_\perp = \pm [(8 c_2^3 \rho_\parallel^3)/(27c_4)]^{1/2}$ between these two regions, which correspond to spinodal lines, are now {\it fold lines} indicating where the LM is folded. The extremizing trajectories appear to be reflected by these lines when they are projected onto the configuration space, so the fold lines may also be called {\it caustics}, in the sense that they form an envelope of the trajectories. Between the two fold lines, the upper (lower) sheet dominates the LDF for $\rho_\perp < 0$ ($\rho_\perp > 0$). The boundary between these two regions coincides with the switching line $\rho_\perp = 0$. The origin $\rho_\parallel = \rho_\perp = 0$, which was previously interpreted as a critical point, coincides with the {\it cusp point} formed by the fold lines.

Note that the middle sheet never plays a dominant role, which has an important implication on the structure of the globally minimizing trajectories: any extremizing trajectory, after being reflected by a fold line, enters the middle sheet and ceases to be dominant. To be more precise, the trajectory is no longer dominant once it crosses the switching line, which occurs always before the reflection at the fold lines. Hence, while the switching line can be identified by tracking the abrupt change of the globally minimizing trajectories, the fold lines are almost never encountered by the typically observable trajectories. This is a notable difference from the caustics appearing in ray optics (in the short wavelength limit) and semiclassical quantum mechanics (in the small $h$ limit), where the caustics are indeed observable patterns formed by extremal trajectories. In these cases, the action has complex values, so the saddle-point method does not impose a minimum criterion in the form of the least action principle \eref{eq:action_lagrangian}. Without such restrictions, nothing prevents the typically observable trajectories from meeting the fold lines~\cite{Dykman1994b}.

Just as the Landau free energy $\cal L$ can have more complicated forms involving more parameters, the LM associated with a Landau-like singularity can have a higher-dimensional structure, which can be obtained from $\cal L$ by a procedure similar to the one discussed above~\cite{Arnold1989}. Even in these cases, multiple extremizing trajectories reaching the same configuration $\brho_\rmf$ requires that the LM has multiple values of $\hbrho_\rmf$ corresponding to the same $\brho_\rmf$. Thus, the folded geometry of the LM is a minimal requirement for a Landau-like singularity. It is known that such a geometry is possible only if the Hamiltonian system is {\it non-integrable} at zero energy, which allows momenta to be multi-valued functions of $\brho$~\cite{Arnold1989}. This reveals an important connection between two seemingly unrelated subjects, namely the existence of a smooth LDF for a weak-noise system and the integrability of the corresponding mechanical system.

\subsubsection{Finite-dimensional models with Landau-like singularities}

Studies of Landau-like singularities in finite-dimensional systems have focused on models with a two-dimensional configuration space. These have the minimum required degrees of freedom for the existence of such singularities. It is known from catastrophe theory that the only structurally stable singularities for such models are cusp points and fold lines; other types of singularities are generically impossible. Thus, the main difference between models does not lie in the types of singularities, but in the global structure formed by the singularities.

The simplest structure involves only a single cusp point and a pair of fold lines. This structure was first identified in a periodically driven nonlinear oscillator by perturbative and numerical methods~\cite{Graham1984a,Graham1984b,Graham1985a,Jauslin1986}, whose Landau-like behaviors are even experimentally observable~\cite{Dykman1996b,Luchinsky1997a,Dykman1998}. See more recent numerical studies~\cite{Bandrivskyy2002,Bandrivskyy2003,Beri2005} for clearer presentations of the structure. For this particular type of systems, it has also been studied whether the driving strength has a nonzero threshold for the onset of singularities~\cite{Kogan2011}. A time-reversed van der Pol oscillator~\cite{Bandrivskyy2003,Beri2005,Maier1996b} and a symmetric double-well system~\cite{Maier1993a,Maier1993b,Maier1996a,Maier2000,Luchinsky1997a,Luchinsky1997b} also have a similar structure of singularities.

A more complicated structure that was identified has the shape of a tree graph: cusp points form the vertices of the tree, and the switching lines become the edges of the tree. A simple tree made by three cusp points and three switching lines was numerically found in \cite{Day1987,Jauslin1987}. Larger trees may contain an infinite number of cusp points and switching lines, as found in some models with a stable limit cycle~\cite{Dykman1996a,Smelyanskiy1997}. See \cite{Smelyanskiy1997} for a general theory of such infinite tree graphs.

\subsection{Other types of singularities}

There are two types of singularities that occur in more restricted circumstances. They do not clearly fall into any of the major categories discussed above.

The first type behaves like a hybrid of the two major categories. It arises when one end of a switching line, on which trajectories from a saddle point and a nearby attractor are equally dominant, coincides with the saddle point itself; see \cite{Dykman1994b} for the conditions for this to happen. Then the saddle point becomes a singularity similar to a cusp point, in the sense that the LDF is trivially differentiable there ($\partial \phi/\partial \brho = \hbrho = 0$ on a limit set, as discussed in \sref{sec:formalism}) and a fold line emanates from it~\cite{Dykman1994b,Chinarov1993,Dykman1994a}. But the singularity does not satisfy the scaling properties expected from a Landau theory, because it is formed by the interference between different limit sets and not by a bifurcation of trajectories. This type of singularities are thus called {\it non-generic cusps}~\cite{Maier1996a}; they have been found in a few systems with a saddle point on the boundary of a basin of attraction, namely a Penning trap~\cite{Dykman1994b,Chinarov1993}, a Selkov model~\cite{Dykman1994a}, and a symmetric double well system without DB~\cite{Hu1988,Maier1993a,Maier1993b,Maier1996a,Maier2000}.

The second type of singularities appear on an attractor and keep the LDF smooth. They are encountered when the attractor itself is about to lose its stability by a bifurcation involving more than one parameter. See \cite{SchimanskyGeier1985,Sulpice1987,Graham1987,Lemarchand1988,Descalzi1989,Tel1989}.

\section{Singularities in driven diffusive systems} \label{sec:inf_dim}

So far we have discussed singularities of the LDFs in finite-dimensional systems. It is natural to ask whether such singularities, or even more complicated ones, are present in spatially extended driven diffusive systems, which have an infinite-dimensional configuration space. For such systems the weak noise limit is natural (see below). In this section, we discuss Landau-like singularities found in these systems, whose origins are essentially the same as those of finite-dimensional cases. To date, these are the only ones identified in driven diffusive systems. We start by showing how the general formalism described in \sref{sec:formalism} can be applied to driven diffusive systems.

\subsection{Large deviations of driven diffusive systems}

We consider a driven diffusive process of a locally conserved quantity $\rho = \rho(x,t)$ in a one-dimensional system with a spatial variable $x \in [0,N]$. The generalization to higher dimensions is trivial. The process satisfies the continuity equation
\begin{equation}\label{eq:continuity}
\partial_t \rho + \partial_x J = 0,
\end{equation}
with the fluctuating current $J$ given by
\begin{equation}\label{eq:current}
J(x,t)=-D(\rho)\partial_{x}\rho+\sigma(\rho)E+\sqrt{\sigma(\rho)}\eta(x,t).
\end{equation}
Here $E$ is a bulk driving force, and $\eta$ is a Gaussian white noise. Using a diffusive rescaling $x \to N^{-1}x$, $t \to N^{-2}t$, and $E \to N^{-1}E$, we find that $\eta$ satisfies
\begin{equation}
\langle \eta(x,t) \rangle = 0, \quad \langle \eta(x,t)\eta(x',t') \rangle = N^{-1}\delta(x-x')\delta(t-t').
\end{equation}
In a thermodynamic limit $N \to \infty$, we naturally obtain a weak-noise problem. Since we now have $x \in [0,1]$, the spatial boundary conditions can be written as
\begin{equation}
\rho(0) = \rho_0, \quad \rho(1) = \rho_1.
\end{equation}
We assume that every part of the system is locally in equilibrium, so that the diffusion coefficient $D$ and the mobility $\sigma$ are related by the local Einstein relation
\begin{equation} \label{eq:einstein}
2D(\rho) = \sigma(\rho)f''(\rho),
\end{equation}
where $f''$ is the second derivative of $f$, the local free energy per unit length, with respect to $\rho$. We also assume that $f$ is a smooth function. By analogy with \eref{eq:ld_principle}, we expect that the steady-state distribution satisfies the large deviation principle
\begin{equation} \label{eq:ld_principle_sp}
P_{\rm s}[\rho(x)] \sim \exp \left\{-N\phi[\rho(x)] \right\},
\end{equation}
where the LDF $\phi$ is now a functional of the spatially extended configuration $\rho(x)$.

As discussed in \sref{sec:formalism}, the LDF $\phi$ can be obtained from a path integral representation of the propagator in a few different ways. For simplicity, we present a method based on the MSR formalism below (see \ref{app:hamiltonian}):
\begin{eqnarray}
\fl P\left[\rho_\rmf(x),t_\rmf|\rho_\rmi(x),t_\rmi\right] &= \int_{\rho(x,t_\rmi)=\rho_\rmi(x)}^{\rho(x,t_\rmf)=\rho_\rmf(x)} {\cal D}[\rho,\eta] \, \delta \left[\dot{\rho} - \partial_x J \right] \exp\left[-N\int_{t_\rmi}^{t_\rmf} \rmd t \int_0^1 \rmd x \, \frac{\eta^2}{2}\right] \nonumber \\
&= \int_{\rho_\rmi(x)}^{\rho_\rmf(x)} {\cal D}[\rho,\hrho,\eta]\, \exp\left[-N\int_{t_\rmi}^{t_\rmf} \rmd t \int_0^1 \rmd x \, \left(\frac{\eta^2}{2}+\hrho\,\partial_t \rho - J\partial_x \hrho\right)\right] \nonumber \\
&= \int_{\rho_\rmi(x)}^{\rho_\rmf(x)} {\cal D}[\rho,\hrho]\, \exp\left[-N\int_{t_\rmi}^{t_\rmf} \rmd t \left( \int_0^1 \rmd x \, \hrho\,\partial_t \rho - H[\rho,\hrho]\right) \right].
\end{eqnarray}
Here $\hrho$ is a momentum field obeying the boundary conditions $\hrho(0) = \hrho(1) = 0$. The functional $H$, defined as
\begin{equation}
\fl H[\rho(x),\hrho(x)] \equiv \int_0^1 \rmd x \, \left[- D(\rho)(\partial_x \hrho)(\partial_x \rho) + \sigma(\rho)E\,\partial_x \hrho + \frac{1}{2}\sigma(\rho) (\partial_x \hrho)^2 \right],
\end{equation}
becomes the Hamiltonian of the system. By the saddle point method, the propagator is approximated as
\begin{equation}
P\left[\rho_\rmf(x),t_\rmf|\rho_\rmi(x),t_\rmi\right] \sim \exp \left\{ -N S\left[\rho_\rmf(x),t_\rmf|\rho_\rmi(x),t_\rmi\right] \right\},
\end{equation}
where $S$ is the minimal action functional which is obtained from
\begin{equation}
S\left[\rho_\rmf(x),t_\rmf|\rho_\rmi(x),t_\rmi\right] = \min_{\rho(x,t),\hrho(x,t)} \int_{t_\rmi}^{t_\rmf} \rmd t \left\{ \int_0^1 \rmd x \, \hrho\, \partial_t \rho - H[\rho,\hrho] \right\}.
\end{equation}
This relation is an analogue of the least action principle stated by \eref{eq:action}; it implies that the propagator is dominated by Hamiltonian trajectories satisfying
\numparts
\begin{eqnarray}
\partial_t \rho &= \frac{\delta H}{\delta \hrho(x)} = \partial_x D(\rho)\partial_x \rho - \partial_x \sigma(\rho)E - \partial_x \sigma(\rho) \partial_x \hrho, \label{eq:hamilton_coords_sp}\\
\partial_t \hrho &= -\frac{\delta H}{\delta \rho(x)} = -D(\rho)\partial_x^2 \hrho - \sigma'(\rho)E\,\partial_x\hrho - \frac{1}{2} \sigma'(\rho) (\partial_x \hrho)^2.
\end{eqnarray}
\endnumparts
Note that, unlike the finite-dimensional cases, the noise effects are now represented by $\partial_x \hrho$ rather than $\hrho$ itself. This is a consequence of the local conservation of $\rho$, which ensures that a divergence operator is always attached to the noise term (see \eref{eq:continuity}). Bearing such subtle differences in mind, we can push the analogy further to obtain the LDF $\phi$ from the action functionals using the Hamiltonian structure described in \sref{sec:formalism}.

It is instructive to first consider systems in equilibrium. In this case, we can easily obtain a DB condition
\begin{equation} \label{eq:ldf_db0_sp}
\frac{\delta \phi[\rho(x)]}{\delta \rho(x)} = \hrho + \hrho^{\rm TR},
\end{equation}
which is an analogue of \eref{eq:ldf_db0}. Besides, in equilibrium the most probable configuration $\bar{\rho}$ satisfies
\begin{equation} \label{eq:most_prob}
-D(\bar{\rho})\partial_x \bar{\rho} + \sigma(\bar{\rho})E = 0,
\end{equation}
which indicates the vanishing average current $\langle J \rangle$. Note that $\hrho^{\rm TR}$ gives rise to the time-reversed equation of motion
\begin{equation}
-\partial_t \rho = \partial_x D(\rho)\partial_x \rho - \partial_x \sigma(\rho)E - \partial_x \sigma(\rho) \partial_x \hrho^{\rm TR},
\end{equation}
which together with \eref{eq:hamilton_coords_sp} implies
\begin{equation} \label{eq:forward_backward}
\sigma(\rho) \, \partial_x (\hrho + \hrho^{\rm TR}) = 2D(\rho) \partial_x \rho - 2 \sigma(\rho) E + {\rm const}.
\end{equation}
Since both sides must be zero for $\rho = \bar{\rho}$, the additive constant is equal to zero. From \eref{eq:einstein}, \eref{eq:ldf_db0_sp}, \eref{eq:most_prob}, and \eref{eq:forward_backward}, we obtain
\begin{equation}
\fl \partial_x \frac{\delta \phi[\rho(x)]}{\delta \rho(x)} = \frac{2D(\rho)}{\sigma(\rho)}\partial_x \rho - 2E = f''(\rho) \partial_x \rho - f''(\bar{\rho})\partial_x \bar{\rho} = \partial_x \left[f'(\rho) - f'(\bar{\rho})\right].
\end{equation}
Using the zero-point prescription $\phi[\bar{\rho}(x)] = 0$, the LDF is obtained as
\begin{equation} \label{eq:ldf_eq_sp}
\phi[\rho(x)] = \int \rmd x \, \left[f(\rho) - f(\bar{\rho}) -f'(\bar{\rho})(\rho-\bar{\rho})\right].
\end{equation}
This shows that the LDF of an equilibrium system is a local functional of the configuration $\rho(x)$, that is, the functional is a sum of local contributions (see also \cite{Bertini2009,Krapivsky2012} for other discussions on this formula). Since the steady-state distribution $P_{\rm s}$ given by \eref{eq:ld_principle_sp} is then a product measure, different parts of the system are uncorrelated. Moreover, \eref{eq:ldf_eq_sp} also shows that $\phi$ is a smooth functional of $\rho(x)$, as $f$ (or equivalently $D$ and $\sigma$) has been assumed to be smooth.

None of these properties are guaranteed if the system is out of equilibrium. It is known that non-equilibrium systems generally develop long-range correlations (see \cite{Dorfman1994} for a review), which manifest themselves in non-local LDFs. Such LDFs were identified in some exactly solvable cases, e.g. the symmetric simple exclusion process (SSEP)~\cite{Bertini2001,Derrida2001,Derrida2002} and the Kipnis--Marchioro--Presutti (KMP) model~\cite{Bertini2005b}. While the LDFs of these examples are still smooth even out of equilibrium, there are other examples in which the LDFs are neither local nor smooth, as discussed below.

\subsection{Landau-like singularities in boundary-driven systems}

While Landau-like singularities were initially identified in a model with a bulk bias~\cite{Bertini2010}, we first discuss those in the absence of any bulk bias (i.e. $E = 0$). In this case the system is kept out of equilibrium by a boundary bias $\rho_0 \neq \rho_1$. Then the behavior of the system is completely determined by $D$ and $\sigma$. The precise condition for $D$ and $\sigma$ to give rise to Landau-like singularities is yet to be clarified, but there are a few boundary-driven models that have been shown to possess these singularities~\cite{Bunin2012b,Bunin2013a}. As a concrete example, we discuss a boundary-driven Ising (BDI) model in some detail.

\begin{figure}[ptb]
\begin{center}
\includegraphics[height=1.3589in,width=3.3167in]{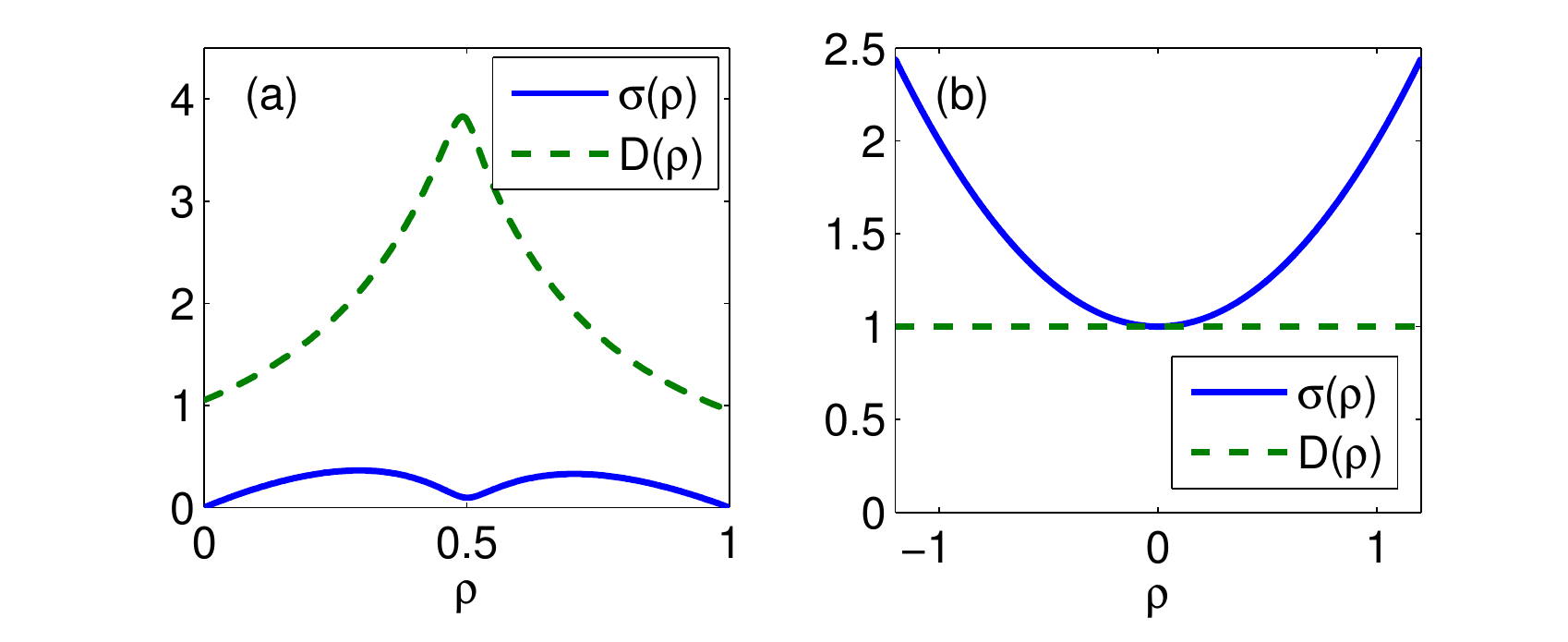}
\caption{The functions $D(\rho)$ and $\sigma(\rho)$ for two boundary-driven models exhibiting Landau-like singularities: (a) the BDI model and (b) the transport model with a quadratic $\sigma$.}
\label{fig:models_sig_d}
\end{center}
\end{figure}

The BDI model is an exclusion process with nearest-neighbor interactions biased towards creation or annihilation of domain walls. Consider a one-dimensional lattice of sites $i=1,..,n$, each of which is either occupied by a particle ($1$) or empty ($0$). Each particle jumps to an adjacent empty site according to the following rates:
\begin{equation}
0100\overset{1+\delta}{\rightarrow}0010,\ 1101\overset{1-\delta}{\rightarrow}1011\ ,1100\overset{1+\varepsilon}{\rightarrow}1010,\ 1010\overset{1-\varepsilon}{\rightarrow}1100,
\end{equation}
with the same rates for the left-right inverted configurations. It is known that these rates lead to an Ising-like steady-state distribution at equilibrium~\cite{Katz1984}. In the continuum limit, the model is described in terms of the particle density profile $\rho(x)$ that evolves according to \eref{eq:continuity} and \eref{eq:current}, with $D$ and $\sigma$ determined by the parameter set $(\varepsilon,\delta)$~\cite{Spohn1991,Katz1984}. Figure~\ref{fig:models_sig_d}(a) shows numerical calculations of these functions for $(\varepsilon,\delta)=\left(0.05,0.995\right)$. Note that $D$ has a peak and $\sigma$ has a local minimum around $\rho=1/2$. A class of boundary-driven models with this type of $D$ and $\sigma$ are called the Katz--Lebowitz--Spohn (KLS) model with zero bulk bias~\cite{Katz1984}. It is easy to check that the model has only one stable configuration $\bar{\rho}$.

\begin{figure}[ptb]
\begin{center}
\includegraphics[height=1.7935in,width=2.7702in]{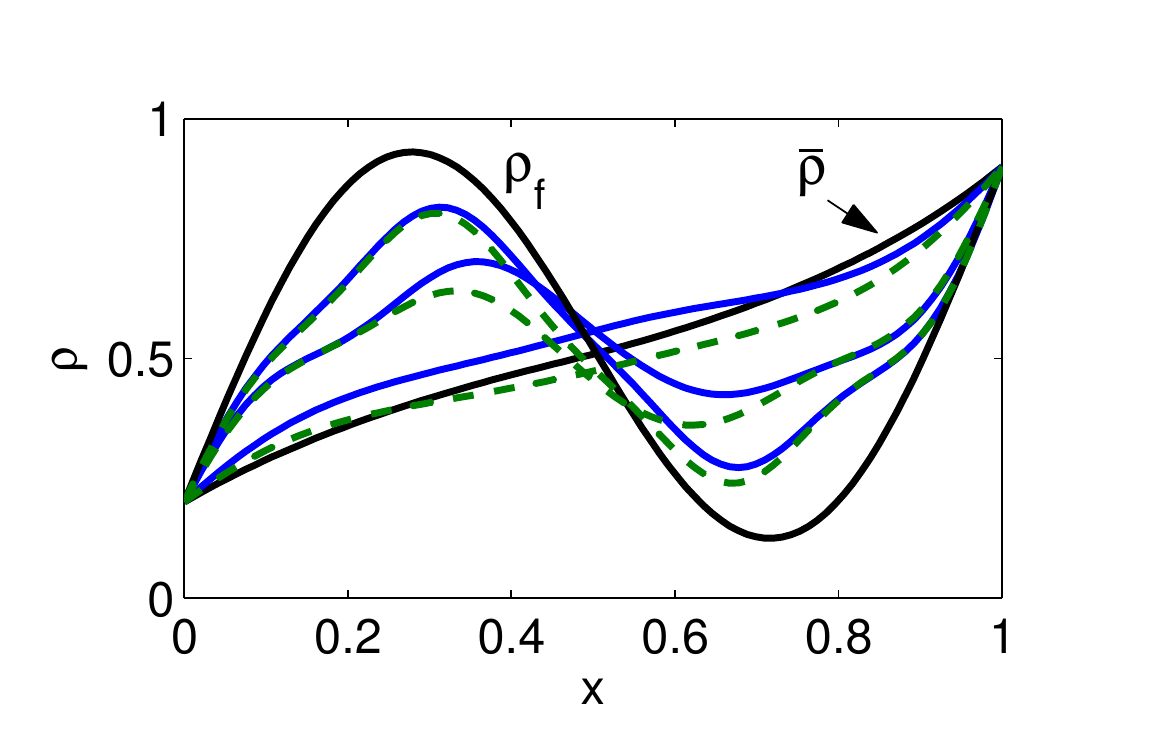}
\caption{Snapshots of two globally minimizing trajectories of the BDI model leading to the same $\rho_{f}$.}
\label{fig:two_paths_bdi}
\end{center}
\end{figure}

\begin{figure}[ptb]
\begin{center}
\includegraphics[height=1.9976in,width=3.9241in]{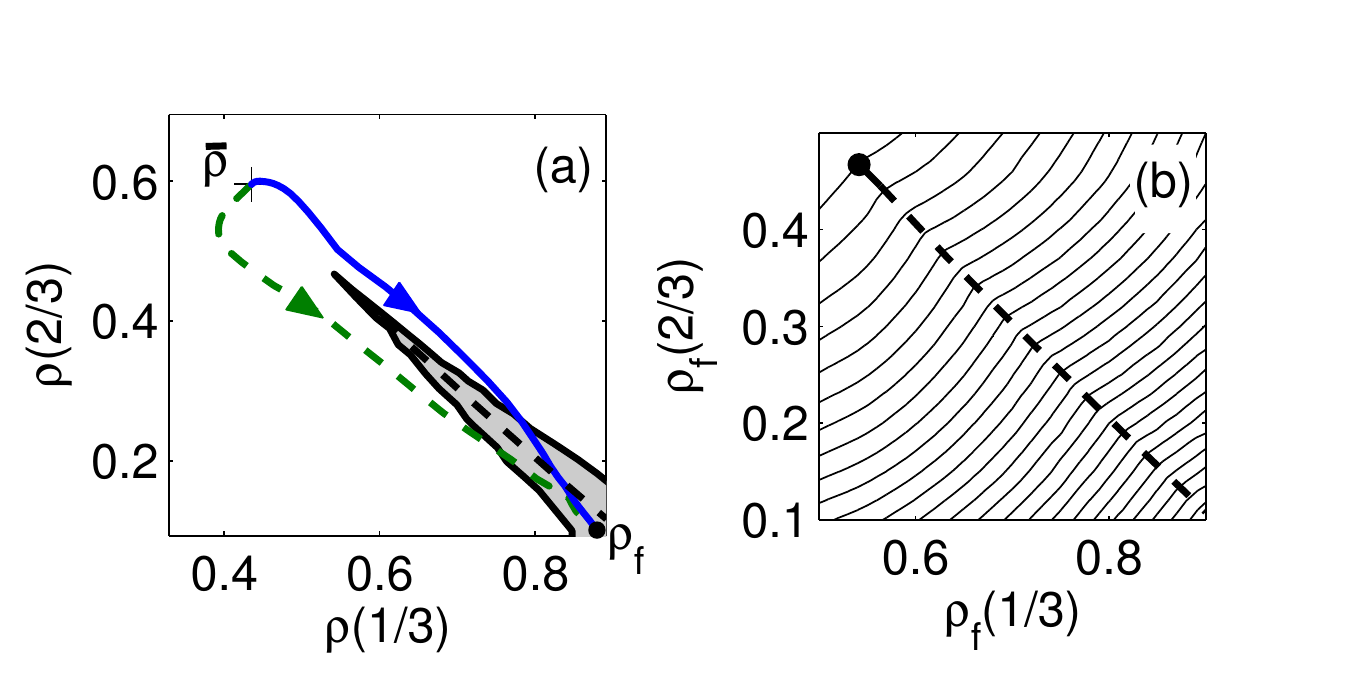}
\caption{A cusp singularity of the BDI model. (a) The cross section at $\rho(1/3)$ and $\rho(2/3)$ of the globally minimizing trajectories shown in \fref{fig:two_paths_bdi}. (b) The switching line (dashed line) with contours of $\phi$ (solid lines).}
\label{fig:bdi_cusp}
\end{center}
\end{figure}

Numerical studies~\cite{Bunin2012b,Bunin2013a} showed that the BDI model has two globally minimizing trajectories for some final configurations, which form a multi-dimensional ``switching surface''. For the convenience of visualization, we focus on the singular behaviors within a two-dimensional subspace of final configurations satisfying
\begin{equation} \label{eq:final_config}
\rho_\rmf(x)  =\bar{\rho}(x)  +\alpha_{1}\sin (\pi x) + \alpha_{2}\sin (2\pi x).
\end{equation}
The subspace is spanned by varying the coefficients $\alpha_1$ and $\alpha_2$, which also have one-to-one correspondence with the configuration components $\rho(x=1/3)$ and $\rho(x=2/3)$. An example of two globally minimizing trajectories is shown in \fref{fig:two_paths_bdi}, in which one trajectory (blue solid lines) starts by raising the total number of particles, while the other (green dashed lines) initially lowers the number. Cross-sections of these same trajectories at $\rho(x=1/3)$ and $\rho(x=2/3)$ are shown in \fref{fig:bdi_cusp}(a), which clearly shows the different routes taken by the trajectories in the configuration space. In agreement with finite-dimensional cases, such trajectories form a pattern in the configuration space associated with a cusp singularity: as shown in \fref{fig:bdi_cusp}(a), there is a region (shaded area) of multiple extremizing trajectories bounded by fold lines (solid lines), and an abrupt change of globally minimizing trajectories occurs across the switching line (dashed line). The contour lines of $\phi$, shown in \fref{fig:bdi_cusp}(b), clearly indicate that $\phi$ is non-differentiable along the switching line.

\begin{figure}
[ptb]
\begin{center}
\includegraphics[
trim=0.000000in 0.008839in 0.000000in -0.008839in,
height=1.4673in,
width=3.509in
]
{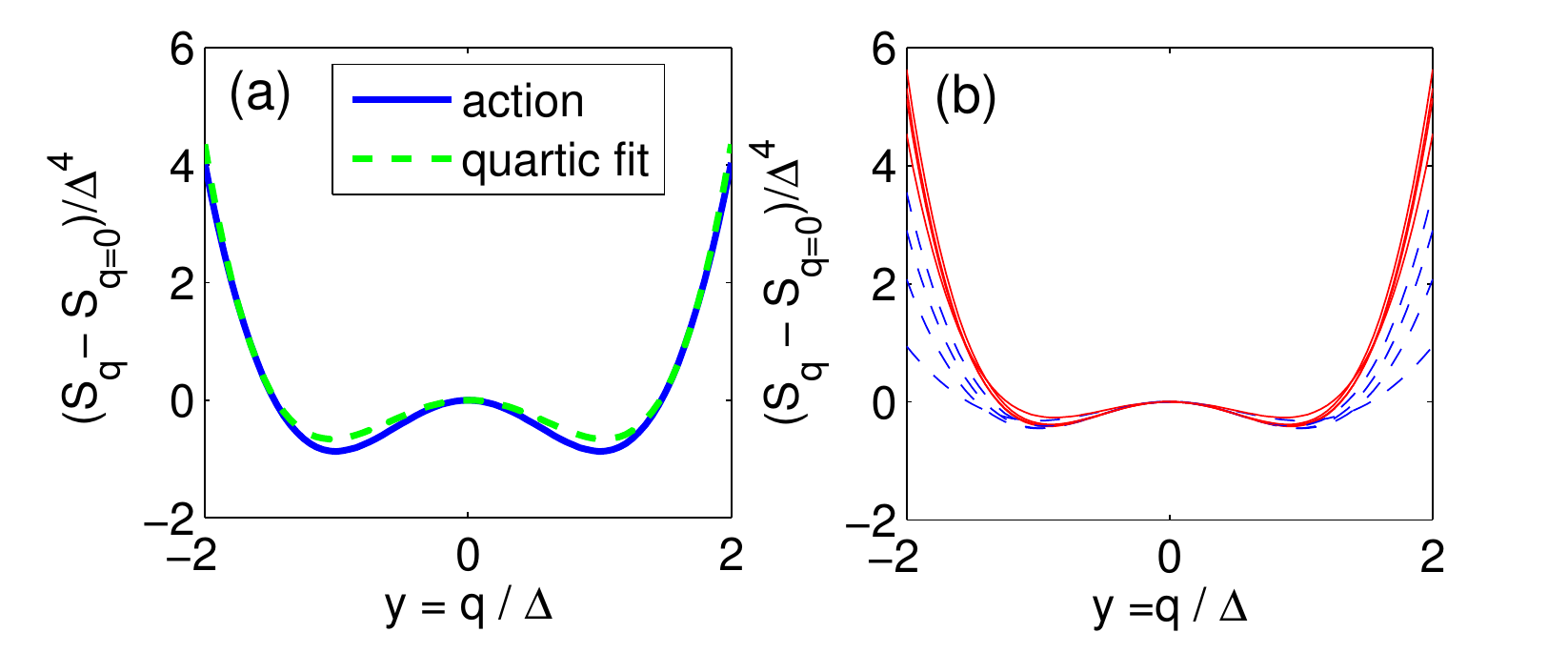}
\caption{(a) The function $(S_{q}-S_{q=0})/\Delta^{4}$ as a function of $y$ at $\rho_\perp = 0$ and $\rho_\parallel=0.012$ (solid line). A deviation is seen from a fit to a quartic function (dashed line). (b) The function $(S_{q}-S_{q=0})/\Delta^{4}$ for different values of $\rho_\parallel$ (dashed lines). Fitting the functions including $y^6$, $y^8$ terms and plotting only the quartic part, the collapse improves significantly (solid lines).}
\label{fig:quartic_polynomials_fit}
\end{center}
\end{figure}

To check whether the cusp singularity is indeed described by a Landau free energy of the form \eref{eq:landau}, one can numerically estimate the action $s_q$ accumulated by a trajectory of the form
\begin{equation}
\rho(x,t) = \rho_{\rm avg}(x,t) + q\,u(x,t),
\end{equation}
where $\rho_{\rm avg}(x,t)$ and $u(x,t)$ are defined from extremal trajectories as in \eref{eq:traj_def}. The final configuration $\rho_\rmf$ is assumed to be on the switching line (i.e. $\rho_\perp = 0$) and very close to the cusp point (i.e. $\rho_\parallel \ll 1$). If $\Delta$ is the value of $q$ minimizing the action and $y \equiv q/\Delta$, we can write
\begin{equation} \label{eq:s_of_y}
s_q-s_{q=0} \sim \Delta^4 \left(\frac{y^4}{4}-\frac{y^2}{2}\right).
\end{equation}
If this scaling form is able to collapse the action functionals estimated at different locations of $\rho_\rmf$, we can conclude that the prefactor of $y^4$ [corresponding to $c_4$ of \eref{eq:landau}] is a finite constant, which confirms the applicability of the Landau free energy \eref{eq:landau}. Figure~\ref{fig:quartic_polynomials_fit}(a) shows that a quartic fit by \eref{eq:s_of_y} approximates the behavior of the action to some extent, albeit with deviations due to higher-order terms. The data collapse is shown in \fref{fig:quartic_polynomials_fit}(b), which indicates that including higher-order terms in the curve fitting (solid lines) improves the data collapse compared to the quartic fits (dashed lines). Thus, we can conclude that the cusp point of the BDI model is indeed described by a Landau free energy \eref{eq:landau}.

Another boundary-driven model, represented by $D(\rho) = 1$ and $\sigma(\rho) = 1+\sigma^2$ [see \fref{fig:models_sig_d}(b)], was found to exhibit a cusp singularity~\cite{Bunin2013a} (note that this model is a spatially extended counterpart of the example discussed in \ref{sssec:qs}). In both boundary-driven models the ratio $\sigma/D$, which is related to the compressibility $\kappa$ by the fluctuation--dissipation relation $\sigma(\rho)/D(\rho) = 2k_{B}T\rho^{2}\kappa(\rho)$, has a local minimum. Numerical results suggest that a sufficiently deep local minimum of $\sigma/D$ may indeed be a condition for the existence of singularities; however, it is still unclear how such a structure of $\sigma/D$ is connected to the form of the action giving rise to trajectory bifurcations~\cite{Bunin2013a}.

We also note that only cusp singularities have been identified in boundary-driven systems. More complicated singularities were found in systems with a bulk bias, which we discuss next.

\subsection{Landau-like singularities in bulk-driven systems}

\begin{figure}
\begin{centering}
\includegraphics[width=0.5\textwidth]{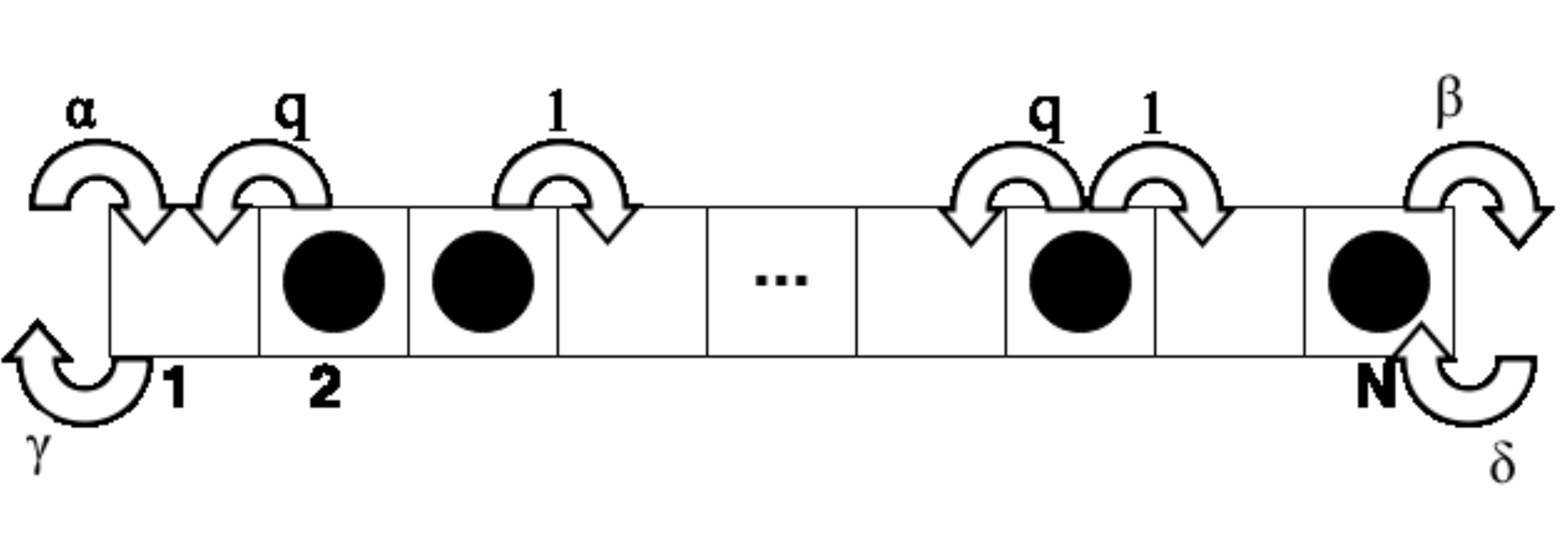} 
\par\end{centering}
\caption{\label{fig:pasep} An illustration of the PASEP. Particles hop one site to the right (left) with rate $1$ ($q$) if the adjacent site is empty. On the left (right) boundary particles enter with rate $\alpha$ ($\delta$) and exit with rate $\gamma$ ($\beta$).}
\end{figure}

The only bulk-driven model that is known to possess a singular LDF is the partially asymmetric simple exclusion process (PASEP). The model is defined on a one-dimensional lattice with $N$ sites (see \fref{fig:pasep}). Each particle on the lattice hops one site to the right with rate $1$ and to the left with rate $q$ provided that the site that it hops to is vacant. For boundary sites $1$ and $N$, a particle enters (exits) the lattice at site $1$ with rate $\alpha$ ($\gamma$) and at site $N$ with rate $\delta$ ($\beta$) if the boundary site is vacant (occupied). 

The continuum limit of the model is described by \eref{eq:continuity} and \eref{eq:current} when $1 - q \sim N^{-1}$, i.e. when the bulk bias decreases with system size. This special case, called the weakly asymmetric simple exclusion process (WASEP), is characterized by the functions~\cite{Spohn1991}
\begin{equation}
E = \frac{N(1-q)}{2}, \quad D(\rho) = 1, \quad \sigma(\rho) = 2\rho(1-\rho)
\end{equation}
and the boundary conditions
\begin{equation}
\rho_0 = \frac{\alpha}{\alpha+\gamma},\quad \rho_1 = \frac{\delta}{\delta+\beta}.
\end{equation}
It can be shown that the model has a unique stable configuration $\bar{\rho}$. Except for the special case when the boundary and bulk biases satisfy the zero-current condition \eref{eq:most_prob}, the model is out of equilibrium. In this case, as discussed above, the LDF is not necessarily smooth. Indeed, if the boundary bias is applied in the direction opposite to the bulk bias ($\rho_0 < \rho_1$ for $E > 0$ and $\rho_0 > \rho_1$ for $E < 0$), the LDF of the model exhibits singularities: cusp singularities are formed as $|E|$ is raised above a threshold value, and an analogue of a tricritical point appears as $|E|$ is increased further. See \cite{Aminov2014} for details.

As a simple example that clearly shows the Landau-like nature of these singularities, in the following we focus on the limit $E \to \infty$, in which case the model becomes equivalent to the ordinary PASEP~\cite{Bertini2010}. The LDF has the form~\cite{Derrida2003,Bertini2010}
\begin{equation} \label{eq:ldf_pasep}
\phi[\rho_\rmf(x)] = \phi_{\rm a}[\rho_\rmf(x)] + \min_{0 < y < 1} {\cal G}[\rho_\rmf(x),y]\ ,\label{eq:PASEP_LDF}
\end{equation}
where $\phi_{\rm a}$ represents the analytic component of $\phi$ and
\begin{eqnarray}
{\cal G}[\rho_{f}(x),y] &= \int_0^y \rmd x \, \left[\rho_\rmf \log (1-\rho_0) + \left(1-\rho_\rmf \right)\log \rho_0\right] \nonumber \\
 & \quad + \int_y^1 \rmd x \, \left[\rho_\rmf \log (1-\rho_1) + (1-\rho_\rmf) \log \rho_1\right].
\end{eqnarray}
It was first shown in \cite{Bertini2010} that there exists $\rho_\rmf$ for which there are two values of $y$ minimizing ${\cal G}$, so that $\phi$ becomes non-differentiable at this $\rho_\rmf$.

\begin{figure}
\begin{center}
\includegraphics[width=0.5\textwidth]{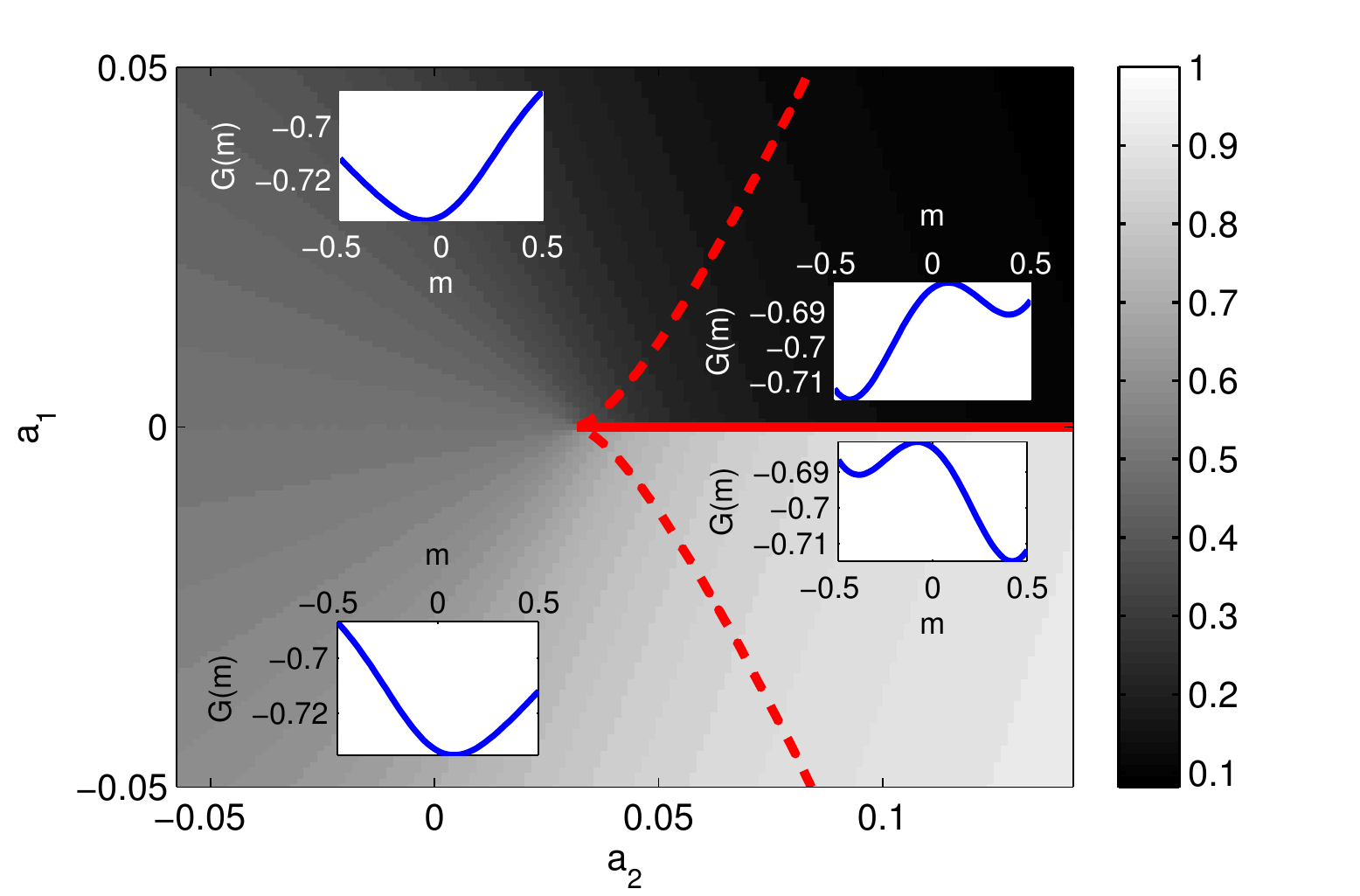}
\end{center}
\caption{\label{fig:full_cusp} The ``phase diagram'' of PASEP for $\rho_\rmf(x)=1/2 - \delta + 2\delta x +a_{1}\sin\left(\pi x\right)+a_{2}\sin\left(2\pi x\right)$ with $\delta=0.1$. The shades indicate the value of minimizing $m$, and the insets illustrate the schematic shapes of $G(m)$ in different areas of the phase space. The dashed line represents the fold lines, and the solid line represents the switching line. All lines meet at the cusp point, an analogue of a critical point.}
\end{figure}

\begin{figure}
\begin{center}
\includegraphics[width=0.5\textwidth]{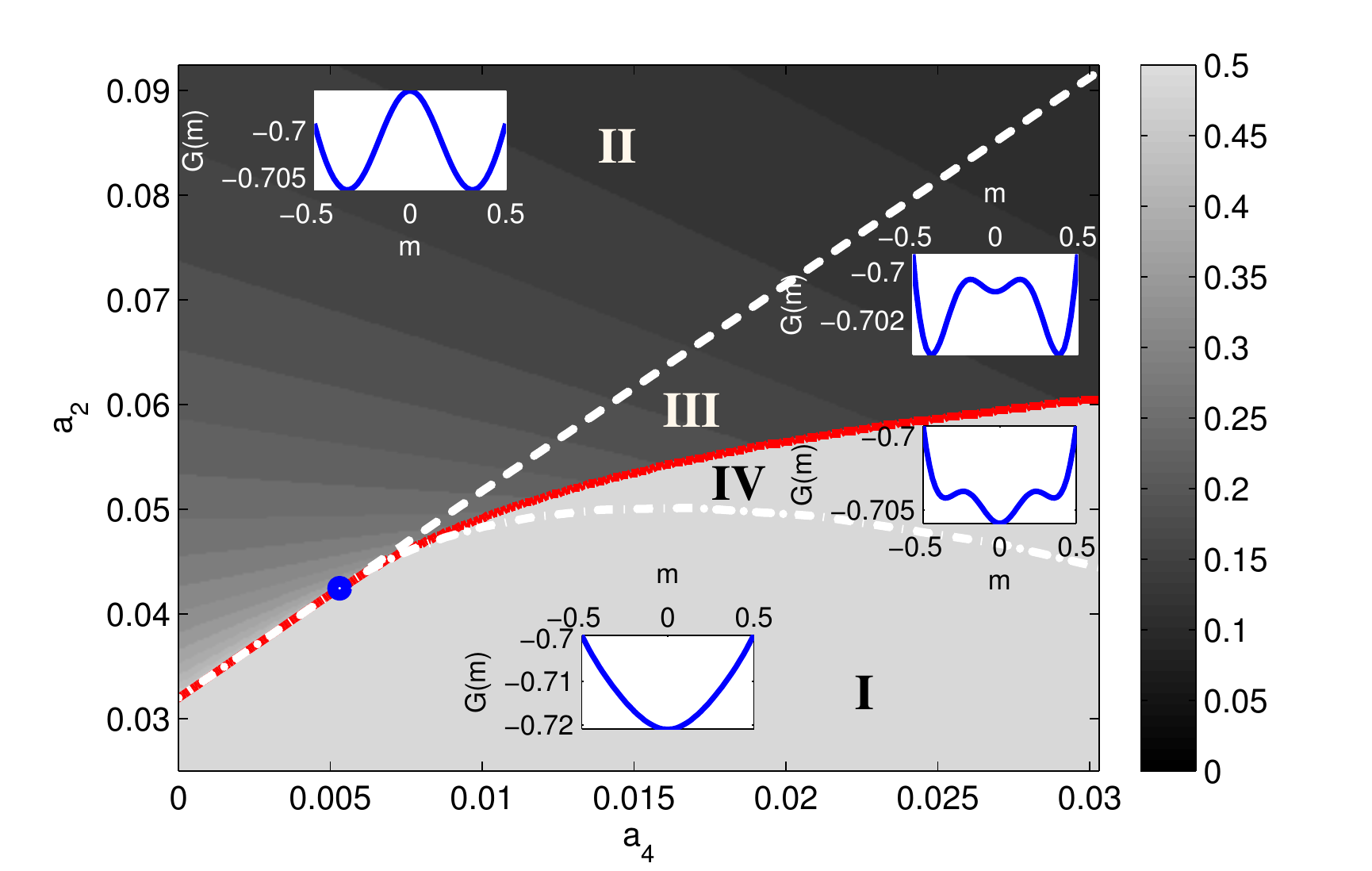}
\end{center}
\caption{\label{fig:full_sym_butterfly} The ``phase diagram'' of PASEP for $\rho_\rmf(x)=1/2 - \delta + 2\delta x +a_{1}\sin\left(\pi x\right)+a_{2}\sin\left(2\pi x\right)+a_{4}\sin\left(4\pi x\right)$ with $\delta=0.1$. The shades indicate the value of minimizing $m$, and the insets illustrate the schematic shapes of $G(m)$ in different areas of the phase space. The solid red line is a switching line, the red--white line a cusp line, and the white lines are fold lines. All lines meet at a tricritical point, marked by a blue dot.}
\end{figure}

An exact Landau theory for this type of singularities can be derived if both the initial configuration (which is stable) and the final configuration have a particle--hole symmetry~\cite{Aminov2014}. The boundary condition $\rho_0 = 1/2 - \delta$ and $\rho_1 = 1/2 + \delta$ with $0 < \delta \le 1/2$ ensures that the stable configuration $\bar{\rho}$ is particle--hole symmetric. The final configuration can be expanded as
\begin{equation} \label{eq:profiles}
\rho_\rmf (x) = \frac{1}{2} - \delta + 2\delta x + \sum_{n=1}^{2n_{\rm max}} a_n \sin (n\pi x),
\end{equation}
where $n_{\rm max}$ is an arbitrary integer. Introducing a change of variable $m=y-1/2$, we can rewrite $\cal G$ as a power series
\begin{equation} \label{eq:G_expansion}
G(m) = 2\log\left(\frac{1+\delta}{1-\delta}\right) \sum_{k=0}^{\infty}c_{k}m^{k}.
\end{equation}
Here $c_0$ is a constant irrelevant to the singular behavior. The other constants (for $k > 1$) are given by
\numparts \label{eq:multicritical_c_all}
\begin{eqnarray}
c_{2k-1} &= \frac{\left(-1\right)^{k-1}}{\left(2k-1\right)!}\sum_{n\in {\rm odd}}\left(-1\right)^{\frac{n-1}{2}}\left(n\pi\right)^{2k-2}a_{n}, \label{eq:multicritical_c_odd} \\
c_{2k} &= \frac{\left(-1\right)^{k+1}}{\left(2k\right)!}\sum_{n\in {\rm even}}\left(-1\right)^{\frac{n}{2}}\left(n\pi\right)^{2k-1}a_{n}. \label{eq:multicritical_c_even}
\end{eqnarray}
\endnumparts
Since $c_1,c_2,\ldots$ are linear combinations of $a_1,a_2,\ldots,a_{2n_{\rm max}}$, we may regard $c_1, c_2, \ldots c_{2n_{\rm max}}$ as free parameters that determine the other coefficients $c_k$ for $k > 2n_{\rm max}$. By choosing an appropriate range of these free parameters, all singular behaviors of the LDF can be captured by the truncated form
\begin{equation}
G(m) \approx 2\log\left(\frac{1+\delta}{1-\delta}\right) \sum_{k=1}^{2n_{\rm max}}c_{k}m^{k}.
\end{equation}
Note that this truncated $G(m)$, with its polynomial form and the requirement of minimization with respect to $m$ [imposed by \eref{eq:ldf_pasep}], is an analogue of a Landau free energy with an order parameter $m$. Depending on the value of $n_{\rm max}$ characterizing $\rho_\rmf$, a Landau free energy for any kind of multicritical point is obtained. For example, if $n_{\rm max} = 1$, we obtain a cusp singularity as illustrated in \fref{fig:full_cusp}. The singular structures around the cusp point are essentially the same as those shown in \fref{fig:lagrangian_manifold}(b). If we choose $n_{\rm max} = 2$ and examine the cross-section $a_1 = 0$, we obtain the structure shown in \fref{fig:full_sym_butterfly} that resembles a phase diagram around a tricritical point: the solid red line, on which the first derivative of the LDF changes discontinuously, meets the red--white line of critical (cusp) points at an analogue of a tricritical point. A pair of spinodal (fold) lines also emanate from the tricritical point. In catastrophe theory, this singular structure is called a symmetry-restricted butterfly catastrophe~\cite{Gilmore1992}.

As discussed in \ref{sssec:landau_lm}, the order parameter $m$ is naturally interpreted as the only momentum component that differs between coexistent globally minimizing trajectories at the final time. We note that the coefficient
\begin{eqnarray}\label{eq:multicritical_c1}
c_1 &= \sum_{n \in {\rm odd}}\left(-1\right)^{\frac{n-1}{2}}a_{n},
\end{eqnarray}
which is obtained from \eref{eq:multicritical_c_odd}, satisfies $\partial \phi/ \partial c_1 \sim m$. This indicates that $m$ is proportional to a momentum $\hat{c}_1$ conjugate to the coordinate $c_1$, which represents the amplitudes of Fourier modes with odd values of $n$. Excitation of such modes indicates that, even if both initial and final configurations are particle--hole symmetric, a globally minimizing trajectory breaks the symmetry. In this sense, $m$ can be regarded as a measure of particle--hole asymmetry. For the simplest case of a cusp singularity ($n_{\rm max} = 1$), we have $c_1 = a_1$, which means that coexistent globally minimizing trajectories evolve differently in terms of the mode $\sin (\pi x)$ associated with $a_1$. This behavior is reflected in the singular structures shown in \fref{fig:full_cusp}, which are created by trajectories that are ``bent'' towards the direction of $a_1$. For other kinds of singularities ($n_{\rm max} > 1$), multiple modes contribute to $m$, and the difference between globally minimizing trajectories are expected to involve more oscillations in $x$ than $\sin (\pi x)$ does; however, this remains yet to be verified.

\section{Summary and outlook} \label{sec:summary}

In this review, we discussed two major scenarios that give rise to singular behaviors in LDFs, their underlying Hamiltonian structure, and the known examples of these behaviors in finite-dimensional systems and spatially extended diffusive systems. The first scenario involves a competition between multiple limit sets and generically produces analogues of discontinuous transitions. The associated type of singularities has been observed only in finite-dimensional systems. The second scenario involves bifurcations of locally minimizing trajectories and is closely related to the integrability of the associated Hamiltonian system. It generates singularities that can be described by Landau theories for critical and multicritical points, which have been observed in both finite-dimensional and driven diffusive systems. For the former only cusp singularities and their tree graphs were observed; this is directly related to the fact that the relevant studies were limited to two-dimensional models. Meanwhile, in the latter case, no analogue of the tree graphs, which were observed in finite-dimensional systems, has been found yet.

As was verified for the Landau-like singularities, it is natural to expect that the origins of singular LDFs are not different for finite-dimensional and driven diffusive systems. Thus, it would be possible to find diffusive systems that exhibit singular behaviors that have been observed only in finite-dimensional cases. In order to achieve the first scenario, one needs to construct driven diffusive systems with metastable states, so that competitions between different limit sets become possible. To produce a graph structure formed by multiple Landau-like singularities, one may consider more sophisticated forms of stable states that correspond to higher-dimensional limit sets such as limit cycles. We also note that some nontrivial singular structures might have been missed by taking cross-sections of the final configuration space for the ease of visualization (see the discussions of \sref{sec:inf_dim}). All of these points indicate that singularities of LDFs in spatially extended systems deserve further investigations.

It would also be interesting to check how the singular structures are affected by relaxing the main assumptions of the general formalism described in \sref{sec:formalism}, such as continuity of the time variable, Gaussian white noise, and arbitrarily small noise intensity. Some studies have already been done in these directions, such as noisy maps with discrete time steps~\cite{Reimann1991}, Langevin equations with weak colored noise~\cite{Einchcomb1995}, and effects of a finite noise amplitude on the globally minimizing trajectories~\cite{Beri2004}. Still, a coherent picture for singular behaviors (or corrections to them) in these more general cases is yet to be established.

Finally, there remains the question of why some non-equilibrium systems have smooth LDFs, even though singular LDFs seem to be more typical. It was pointed out that some non-equilibrium systems with exactly known LDFs (which are also smooth) satisfy hidden DB relations, which are revealed only by non-local mappings to equilibrium systems~\cite{Tailleur2007,Tailleur2008}. Since DB guarantees smooth LDFs for equilibrium diffusive systems, one might guess that the presence of hidden DB and properties of the associated mapping are deeply related to the smoothness of LDFs in non-equilibrium systems.

{\bf Acknowledments} YB and YK are supported by the Israeli Science Foundation. YK is grateful for his collaborators on related topics over the years Avi Aminov, Guy Bunin, Vivien Lecomte and Daniel Podolsky. YK also thanks the Galileo Galilei Institute for Theoretical Physics for hospitality.

\appendix

\section{Other derivations of Hamiltonian mechanics}
\label{app:hamiltonian}
We note that \eref{eq:action} can be derived more directly from \eref{eq:path_int} using the standard Martin--Siggia--Rose (MSR) procedure~\cite{Martin1973,DeDominicis1976,Janssen1976,DeDominicis1978}, which introduces $\hbrho$ by the Fourier transform
\begin{equation} \label{eq:transform}
\delta (\dot{\brho} - \bi{F} - G \bfeta) = \int_{-\rmi \infty}^{\rmi \infty} {\cal D}\hbrho \, \exp \left[ -N\int_{t_\rmi}^{t_\rmf} \rmd t\,\hbrho \cdot ( \dot{\brho} - \bi{F} - G\bfeta ) \right].
\end{equation}

Here the unit of $\hbrho$ has been chosen so that the exponent in ({\ref{eq:transform}) conveniently scale as $N$. Then \eref{eq:path_int} can be written as
\begin{eqnarray}
\fl P(\brho_\rmf,t_\rmf|\brho_\rmi,t_\rmi) &= \int_{\brho(t_\rmi)=\brho_\rmi}^{\brho(t_\rmf)=\brho_\rmf}{\cal D}[\brho,\hbrho,\bfeta] \, \exp \left\{ -N \int_{t_\rmi}^{t_\rmf} \rmd t \, \left[ \frac{\bfeta^2}{2} + \hbrho \cdot ( \dot{\brho} - \bi{F} - G\bfeta )\right] \right\} \\
\fl &= \int_{\brho(t_\rmi)=\brho_\rmi}^{\brho(t_\rmf)=\brho_\rmf}{\cal D}[\brho, \hbrho] \, \exp \left\{ -N \int_{t_\rmi}^{t_\rmf} \rmd t \, \left[ \hbrho \cdot \dot{\brho} - H(\brho,\hbrho) \right] \right\} \label{eq:propagator_hamiltonian}.
\end{eqnarray}
While $\hbrho$ is integrated along the imaginary axes up to this point, we take the Wick rotation $\hbrho \to -\rmi\hbrho$ to place $\hbrho$ on the real axes. After then, the saddle-point method can be applied to reproduce \eref{eq:action}.

\section*{References}
\bibliographystyle{iopart-num}
\bibliography{JSTAT2015-BK}

\end{document}